\documentclass[aps,pra,12pt,reprint,superscriptaddress,citeautoscript,floatfix]{revtex4-1}
\usepackage{amsmath}
\usepackage{graphicx}
\usepackage{epstopdf}
\usepackage{hyperref}
\usepackage{subfigure}

\let\oldhat\hat
\renewcommand{\vec}[1]{\mathbf{#1}}
\renewcommand{\hat}[1]{\oldhat{\mathbf{#1}}}
\newcommand{\upperRomannumeral}[1]{\uppercase\expandafter{\romannumeral#1}}

\begin{document}


\title{Imaging the paramagnetic nonlinear Meissner effect in nodal gap superconductors}

\author{Alexander P. Zhuravel}
\affiliation{B. Verkin Institute for Low Temperature Physics and Engineering, Kharkov  61103, Ukraine}

\author{Seokjin Bae}
\affiliation{Center for Nanophysics and Advanced Materials, Department of Physics, University of Maryland, College Park, MD 20742, USA}

\author{Sergey N. Shevchenko}
\affiliation{B. Verkin Institute for Low Temperature Physics and Engineering, Kharkov  61103, Ukraine}
\affiliation{V. Karazin Kharkov National University, Kharkov 61022, Ukraine}

\author{Alexander N. Omelyanchouk}
\affiliation{B. Verkin Institute for Low Temperature Physics and Engineering, Kharkov  61103, Ukraine}

\author{Alexander V. Lukashenko}
\affiliation{Physikalisches Institut, Karlsruhe Institute of Technology, 76131 Karlsruhe, Germany}

\author{Alexey V. Ustinov}
\affiliation{Physikalisches Institut, Karlsruhe Institute of Technology, 76131 Karlsruhe, Germany}
\affiliation{Russian Quantum Center, National University of Science and Technology MISIS, Moscow 119049, Russia}

\author{Steven M. Anlage}
\affiliation{Center for Nanophysics and Advanced Materials, Department of Physics, University of Maryland, College Park, MD 20742, USA}

\date{\today}

\begin{abstract}
Boundary surfaces of nodal gap superconductors can host Andreev bound states (ABS) which develop a paramagnetic response under external RF field in contrast to the bulk diamagnetic response of the bulk superconductor. At low temperature this surface paramagnetic response dominates and enhances the nonlinear RF response of the sample. With a recently developed photoresponse imaging technique, the anisotropy of this ``paramagnetic" nonlinear Meissner response, and its current direction (angular) and RF power dependence has been systematically studied. A theoretical model describing the current flow in the surface paramagnetic Andreev bound state, the bulk diamagnetic Meissner state, and their response to optical illumination is proposed and it shows good agreement with the experimental results.
 
\end{abstract}


\maketitle

\section{Introduction}
\par The spontaenous expulsion of magnetic flux from the bulk of a superconductor is known as the Meissner effect. In the presence of a weak (both DC and RF) field, the applied field is screened by super-current flow with a density $\vec{j}_s=-en_s \vec{v}_s$ that is proportional to the velocity $\vec{v}_s$ of the condensate. The thickness of the screening surface layer is on the order of a temperature dependent magnetic penetration depth, $\lambda(T)$. At higher field, the super-fluid density $n_s$ becomes dependent on $\vec{v}_s$ (for $\vec{v}_s$ comparable to the critical depairing velocity $\vec{v}_{dp} = \hbar/m^*\xi$) due to Cooper pair breaking. Here $\xi$ is the BCS coherence length and $m^*$ is the effective mass of Cooper pairs. This in turn leads to a field and current dependent magnetic penetration depth, resulting in the nonlinear Meissner effect (NLME).\cite{Yip1992,DXu1995,Groll2010,Gittleman1965PR}

\par The NLME is sensitive to intrinsic properties of a superconducting material including the underlying pairing symmetry. For example, cuprate superconductors with $d_{x^2-y^2}$ gap symmetry of the order parameter are expected to have a strong NLME at temperatures $T\rightarrow 0$, due to the low-lying excitations along the superconducting gap nodal lines.\cite{Yip1992} The $d_{x^2-y^2}$ pairing state also leads to an angular dependent nonlinear response for fields in the $ab$-plane depending on current flow relative to the locations of gap nodes on the Fermi surface.\cite{DXu1995} This (local) anisotropic NLME (aNLME) was initially predicted as a linear magnetic field dependence of the magnetic penetration depth at low temperatures with $1/\sqrt{2}$ anisotropy at $T=0$.\cite{DXu1995} Later, the theories were generalized to all temperatures in terms of nonlinear microwave intermodulation response of a nodal superconductor and a practical method for probing NLME and its $ab$-plane anisotropy was worked out.\cite{Dahm1996,Dahm1999,Dahm1997} The nonlinear superfluid density, $n_s(T,\vec{j}_s)=n_s(T)[1-b_\chi (T)(j_s/j_c)^2]$ becomes dependent not only on $T$ and $\vec{j}_s$, but also on the angle $\chi$ between supercurrent density and directions of the superconducting gap antinodes (which is equivalent to $a$- or $b$-axis direction in the case of a $c$-axis oriented epitaxially grown YBa$_2$Cu$_3$O$_{7-x}$ (YBCO) film). Here, $b_\chi$ is the angular dependent nonlinear Meissner coefficient demonstrating nodal magnitude correction $b_N(\chi=\pi/4)$ almost two times higher than anti-nodal one $b_{AN}(\chi=0)$ at lower reduced temperatures.\cite{Dahm1996} It was found that the anisotropy in the NLME of cuprate high-$T_c$ superconductors (HTS) is weak at high temperatures, and only becomes significant for $T/T_c < 0.6$.\cite{Dahm1996} In addition, it was shown that $b_N$ is expected to grow as $1/T$ for $T/T_c < 0.2$,\cite{Dahm1997} before crossing over to another temperature dependence, depending on the purity of the material.\cite{DXu1995,Li1998,Dahm1999}

\par Many experimental efforts have been made to observe the NLME in $d_{x^2-y^2}$ superconductors.\cite{Bidinosti1999,Bhattacharya1999,Carrington2001,Oates2004,Maeda1995,Carrington1999} The first indirect confirmation of the existence of gap nodes in single crystals of the Bi$_2$Sr$_2$CaCu$_2$O$_{8-x}$ (Bi-2212) system has been demonstrated by Maeda et al., showing linear behavior of $\Delta\lambda(H,T)$ on dc magnetic field H.\cite{Maeda1995} In subsequent experiments on detection of the NLME in cuprates through transverse magnetization\cite{Bhattacharya1999} and magnetic penetration depth,\cite{Bidinosti1999,Carrington2001} the results have been inconclusive as well most likely because of a very small field range of the Meissner state. This is argued by the fact that the NLME becomes significant only in fields H of the order of the thermodynamic critical field $H_c>H_{c1}$ masking nodal quasiparticle excitation at sufficiently strong rf field by other stronger nonlinear effects, such as vortex penetration at fields above the lower-critical field $H_{c1}$.\cite{Groll2010} In addition, the NLME is very small and tends to be obscured by extrinsic effects and thus the manifestation of NLME becomes dependent on the sample and the sensitivity of the measuring technique. Later, the first experimental evidence of the existence of the NLME in high-temperature superconducting YBCO was clearly demonstrated\cite{Oates2004,Benz2001,Leong2005} using the sensitive nonlinear microwave measurement technique of intermodulation product distortion. However, it remained unclear whether the expected anisotropy could be demonstrated to establish experimental verification of the NLME. The best way to elucidate this issue is through a spatially resolved imaging technique. A series of sensitive nonlinear near-field microwave microscopes have been developed to image local sources of nonlinear electrodynamic response in superconductors.\cite{SCLee2003APL,SCLee2003IEEE,SCLee2005PRB,SCLee2005PRB2,Mircea2009PRB,Tai2013IEEE,Tai2014APL,Tai2015PRB,Tai2017PhysicaC} However, these microscopes are not well suited for anisotropy studies. One can examine the nonlinear Meissner effect uniquely in terms of the nodal directions by exploiting special orientations for the current flow, as has been shown in our previous work.\cite{Zhuravel2013} 

\par An additional contribution to the NLME anisotropy of HTS arises from Andreev bound states (ABS)\cite{Hu1994} as a result of participation of, for example, the (110)-oriented surface of a $d_{x^2-y^2}$ superconductor. The sign change of the order parameter at the gap nodes causes an incoming quasiparticle to experience a strong Andreev reflection at the surface. A bound state results from the constructive interference of electron-like and hole-like excitations which originate from such a reflection.\cite{Aprili1999} These states give rise to a paramagnetic contribution to the screening.\cite{Fogelstrom1997}

\par This paramagnetic Meissner effect was studied theoretically\cite{Fogelstrom1997,Higashitani1997} and experimentally.\cite{Braunisch1992,Walter1998PRL,Geim1998,Barbara1999PRB,Ilichev2003,Li2003} For cuprates, (110) interfaces also occur at twin boundaries, which are formed spontaneously during epitaxial film growth. The NLME associated with ABS has been established by tunneling,\cite{Aprili1999} and penetration depth measurements,\cite{Carrington2001,Walter1998PRL} for example. Theory by Barash, Kalenkov, and Kurkijarvi\cite{Barash2000} and Zare, Dahm, and Schopohl\cite{Zare2010PRL} predicts an aNLME associated with ABS having a strong temperature dependence at low temperatures, eventually dominating that due to nodal excitations from the bulk Meissner state. 

\par In what follows, we will refer to the diamagnetic current as the \textit{Meissner} or \textit{bulk} current, while the current flowing next to the boundary and related to ABS will be referred to as the \textit{surface} or \textit{ABS} current. It is thought that weak \textit{bulk} currents give rise to a monotonically decreasing value of the penetration depth as the HTS film is cooled down. On the other hand, the \textit{surface} quasiparticle flow from the ABS enhances the local field and serves to effectively increase the penetration depth. The total effect leads to the appearance of a local minimum in the effective penetration depth as a function of temperature. The predicted penetration depth crossover temperature for a typical cuprate superconductor like YBCO is $T_m=T_c/\sqrt{\kappa}\sim 10$ K,\cite{Zare2010PRL} assumes no impurity scattering, where $\kappa=\lambda_0/\xi_0$ is the Ginzburg-Landau parameter of the superconductor and $\xi_0=hv_F/\pi\Delta_0$ is the coherence length. 

\par One can speculate in this case that the low-temperature NLME should be associated mainly with the ABS contribution. This, in turn, calls for further investigation of the inductive/dissipative origin of the NLME from the boundary surface assuming the presence of a nonlinear surface conductivity associated with qusiparticle flow in the thin surface layer of thickness $\sim\xi_0$. However, it is undeservedly ignored in almost all research of the NLME which is known to us. Here, we propose a new method to quantitatively measure and image the aNLME from ABS of a superconductor. This experiment reveals signatures of the nodal structure of the sample using a procedure of local (resistive and inductive) nonlinear response partition combined with laser scanning microscopy.

\par It was demonstrated recently\cite{Zhuravel2013,Zhuravel2012PRB} that the observation of the photoresponse (PR) allows direct visualization of the anisotropy of the nonlinear Meissner effect. In this paper we will present further experimental evidence for the strong anisotropic response of $d_{x^2-y^2}$ superconducting films in Sec. \ref{Experiment} especially focusing on that from surface ABS. Following in Sec. \ref{Theory}, we will provide a microscopic model which describes quasiparticle flow in the surface ABS in terms of various experimental parameters and its mechanism to give paramagnetic nonlinear Meissner response. Then the calculated response from the theory will be compared to that of the experimental data in Sec. \ref{Comparison of Data and Theory} where it turns out that they show good agreement. 

\begin{figure}
		\includegraphics[width=0.7\columnwidth]{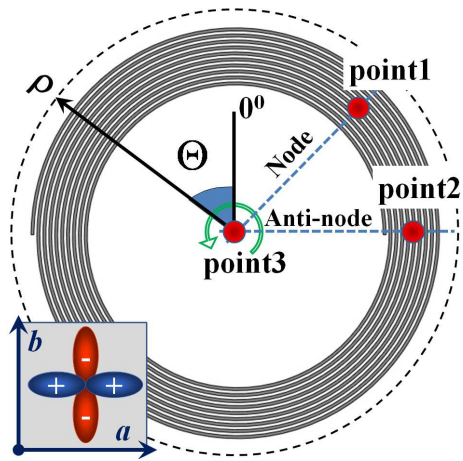}
		\caption{\label{fig:sprial_shape} Schematic sketch of spiral geometry, definition of radial ($\rho$) and angular ($\Theta$) coordinates, and directions of the crystallographic $a$, $b$ axes, along with the orientation of the $d_{x^2-y^2}$ gap in YBCO. Red points 1, 2 and 3 indicate positions utilized for local LSM PR measurements }
\end{figure}

\section{Experiment}\label{Experiment}
\subsection{Self-resonant superconducting sample}

\par The examined sample was a self-resonant superconducting structure based on a thin film spiral geometry. It is manufactured from a $c$-axis normal oriented superconducting YBCO films epitaxially deposited to a thickness of 300 nm by thermal co-evaporation onto an 350 $\mu$m thick single crystal $MgO$ ($\epsilon_r\sim9.7$) substrate.\cite{Ghamsari2013APL} The HTS film is patterned subsequently into a spiral resonator by contact photolithography and wet chemical etching. The spiral has an inner diameter of $D_i=4.4$ mm, an outer diameter of $D_o=6$ mm, and consists of $N = 40.5$ turns of about $s = 10$ $\mu$m width YBCO stripe with $c = 10$ $\mu$m gap between stripes, winding continuously from the inner to outer radii with Archimedean shape (see the schematic diagram in Fig. \ref{fig:sprial_shape}). The same sample configuration was used previously for LSM imaging of the temperature dependent aNLME through the nonlinear electrodynamic response of both (bulk) gap nodes and (surface) Andreev bound states.\cite{Zhuravel2013} A set of such resonators was fabricated at the University of Maryland (College Park, USA).\cite{Ghamsari2013IEEE} The spiral was originally proposed as a compact magnetic meta-atom for use in superconducting metamaterials with a deep sub-wavelength physical dimension of $\lambda_r/D_0\sim 1000$, where $\lambda_r$ is the free-space wavelength at its fundamental resonance.\cite{Anlage2011JOptics,Kurter2010} Previous LSM measurements of superconducting spirals have revealed ``hot spot'' formation at high driving RF powers.\cite{Kurter2011PRB} Here, we give an example of LSM characterization of the resonator at the third harmonic frequency of about 257 MHz where it demonstrates a loaded $Q_L\sim650$ at $T=4.8$ K. From the series of previously tested samples we chose one that is characterized by the maximal ``penetration depth crossover temperature" ($T_m = 7.3$ K) that separates the temperature regimes of bulk NLME and ABS NLME responses. This allowed us to carry out almost all of the following measurements in a convenient operating temperature range $T>4.2$ K.

\par There are a few more unique properties of the studied resonant spiral. First, the distribution of standing wave currents on the spiral are well approximated as those of a one-dimensional transmission line resonator that is rolled into a spiral, as verified by detailed LSM imaging.\cite{Kurter2011PRB} Second, the shape of the $n$-th mode standing wave pattern can be modeled (using polar coordinates $\rho$, $\Theta$ of Fig. \ref{fig:sprial_shape}) as $j_{RF}(n,\rho)\simeq j_0\sin\left(n\pi\left(2\rho/D_0\right)^2\right)$, showing independence of radially $\rho$-averaged currents on angular position $\Theta$, where $j_0$ is the peak value of total RF current $j_{RF}=j_s-j_{qp}$, and $j_{qp}$ is the quasiparticles backflow.\cite{Maleeva2015} Third, the RF currents (at least in the low-order modes) circle the spiral almost 40 times, repeatedly sampling all the angular directions of current flow relative to the planar Cu–O bonds i.e. all parts of the in-plane Fermi surface.\cite{Zhuravel2012PRB} And finally, since the direction of the current is tangential to the spiral, the angular position ($\Theta$ in Fig. \ref{fig:sprial_shape}) of the spiral in real space has a one-to-one mapping relation to each direction $\chi$ in momentum space. As an example, for the $d_{x^2-y^2}$ gap $\Delta(\chi)=\Delta_0(T,j)\cos(2\chi)$, the gap antinodal direction ($k_x$, $k_y$) corresponds to the (100) or (010) direction ($\Theta=0^\circ,90^\circ,180^\circ,270^\circ$) in real space, and the gap nodal ($k_{xy}$) direction corresponds to the ($\pm$110) direction ($\Theta=45^\circ,135^\circ,225^\circ,315^\circ$). Therefore, the method of laser scanning microscopy (LSM) can be used to locate the positions of nodal directions directly in real space coordinates using the advantages of the proposed sample.    

\begin{figure}
		\includegraphics[width=1\columnwidth]{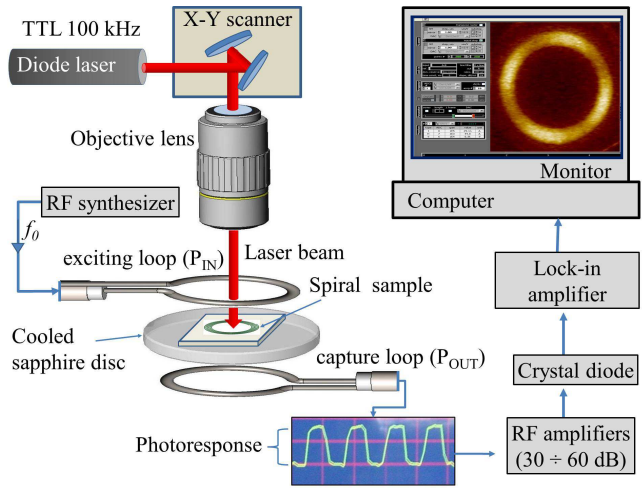}
		\caption{\label{fig:LSM_setup} Schematic diagram of LSM optics and the microwave electronics used for 2D visualization of anisotropic NLME. A single YBCO spiral is sandwiched between two magnetic loops extended from coaxial cables. The red line shows the flow of x-y scanning laser beam through the optical train while blue arrowed line shows the path of injected and transmitted microwave signals. Inset in monitor shows typical screen shot of visualizing software. Bottom inset illustrates amplitude of TTL modulated photoresponse in the form of a measured oscilloscope signal. }
\end{figure}

\subsection{Global transmission data}
\par To obtain the a global microwave response of the spiral, the RF transmission coefficient $S_{21}(f)$ measurements are carried out using a Microwave Vector Network Analyzer (Anritsu MS4640A) that is SMA coupled by stainless semi-rigid coaxial cables to two loop antennas placed inside an optical cryostat. The sample is centered between these circular loops of RF magnetic field probes, 6 mm in inner diameter, whose planes are positioned parallel above/below the sandwiched YBCO spiral structure as shown in Fig. \ref{fig:LSM_setup}. For reliable cooling in vacuum, the back side of the MgO substrate is glued by cryogenic grease to a sapphire disc that is supported on a copper holder which controls temperature of the sample between 100 K and 2.5 K with an accuracy of 1 mK. Excitation of the HTS spiral at different microwave power levels $P_{RF}$ between -30 dBm and +10 dBm is provided by the top loop while the bottom one plays a role of a transmission pick-up probe. More details about the measurement setup can be found elsewhere.\cite{Zhuravel2013,Ghamsari2013APL,Ghamsari2013IEEE} 

\begin{figure}
		\includegraphics[width=0.8\columnwidth]{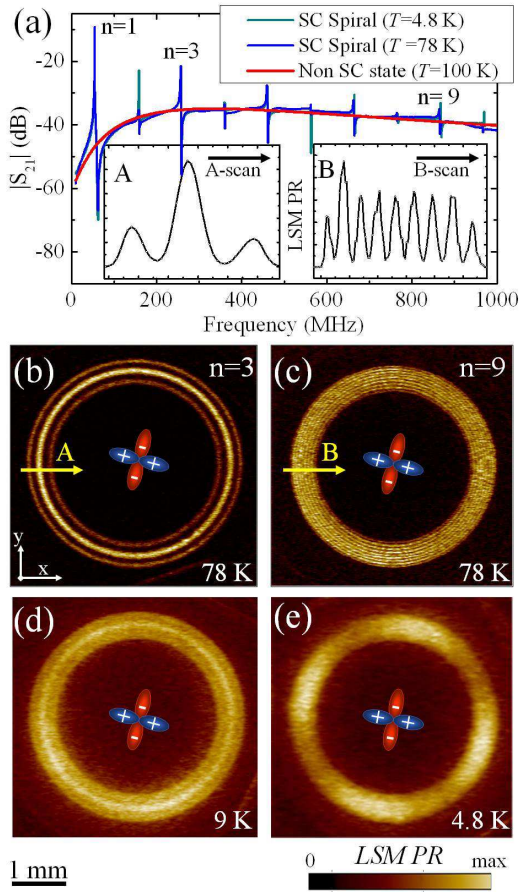}
		\caption{\label{fig:transmission_profile} (a) Transmission coefficient $|S_{21}|$ vs. frequency on the YBCO/MgO spiral showing the fundamental (n=1) and higher harmonic resonances at $P_{RF}=-21$ dBm and $T=$ 4.8 K (green), 78 K (blue), 100 K (red); LSM photoresponse (PR) images of the same spiral showing RF current distribution in the spiral corresponding to (b) the 3-rd and (c) the 9-th resonant modes in the transmission data at 78 K. Insets A and B in part (a) show profiles of $PR(x)$ distribution along corresponding radial line cuts as outlined by arrows A in part (b) and B in part (c). Images (d) and (e) demonstrate bulk diamagnetic aNLME PR and surface paramagnetic ABS PR in the 3-rd resonant mode above and below $T_m\sim$7.3 K, respectively. }
\end{figure}

\par Fig. \ref{fig:transmission_profile}(a) shows the global spectrum of transmission scattering characteristics $|S_{21}(f)|$ of the YBCO/MgO spiral resonator measured at three different temperatures at $P_{RF}=-21$ dBm. The reference (red solid line) transmission spectrum is taken in the normal (non-superconducting) state of the spiral demonstrating dissipative suppression of the all RF resonances at temperature $T=$ 100 K well above $T_c$ of YBCO. Transmission data of the same spiral at 78 K [blue curve in Fig. \ref{fig:transmission_profile}(a)] describes the response of the linear Meissner phase at $P_{RF}=-21$ dBm. Ten almost equidistantly distributed resonances are clearly visible.\cite{Ghamsari2013IEEE} As seen from this data, the frequency $f_1$ of the fundamental harmonic is as low as 74 MHz, followed by higher modes $f_n\simeq nf_1$, where $n = 1, 2,\cdots, N$ is the resonant mode number. The photoresponse (PR) which is a quantity proportional to $j^2_{RF}(x,y)$\cite{Culbertson1998} in the spiral under these circumstances was imaged by using the LSM technique as in Fig. \ref{fig:transmission_profile}(b)-(c). More details on the LSM PR method will be discussed in Sec. \ref{Spatially resolved results}. The LSM PR image of the YBCO spiral near the third resonance tone clearly shows three concentric circles of the standing-wave pattern in Fig. \ref{fig:transmission_profile}(b) as expected. The brightest areas here correspond to peak values of the currents flowing along the windings while zero current density looks black. The 9-th harmonic [Fig. \ref{fig:transmission_profile}(c)] shows nine large-amplitude circles, suggesting that the behavior of the spiral below $T_c$ is described well by TEM modes similar to ones in a linear strip-line resonator where the number of the half-wave standing wave patterns of the $j_{RF}$ distribution is equal to the corresponding $n$ number.\cite{Kurter2011IEEE,Hooker2013,Maleeva2014} One can emphasize that the distribution of $j_{RF}^2(x,y)$ at 78 K is isotropic relative to the superconducting gap configuration of YBCO, as shown schematically in the center of the LSM PR images.

\par Almost the same resonant spectrum of $|S_{21}(f)|$ is observed at decreasing temperature down to 4.8 K and the same $P_{RF}=-21$ dBm (see green curve in Fig. \ref{fig:transmission_profile}(a)). At the same time, the PR is considerably degraded due to a small temperature dependence of the magnetic penetration depth which stays at an almost fixed value below $T/T_c < 0.5$. At significantly lower temperature $T/T_c<0.2$, however, the LSM PR arises again as another form of anisotropic image demonstrating the nonlinear electrodynamic response of both (bulk) gap nodes [Fig. \ref{fig:transmission_profile}(d)] and (surface) Andreev bound states [Fig. \ref{fig:transmission_profile}(e)] despite the unchanged shape of $|S_{21}(f)|$.\cite{Zhuravel2013} Since the surface paramagnetic current shows a sharp increase at low temperature ($T/T_c<0.1$) as will be shown in the theory section, one can expect that the LSM PR below $T_m \sim 7.3$ K arises largely from the anisotropic ABS response. However, this fact is in no way indicated by the behavior of the globally measured $|S_{21}(f)|$, and will be the subject of the remainder of this paper.

\par
\begin{figure}
		\includegraphics[width=1\columnwidth]{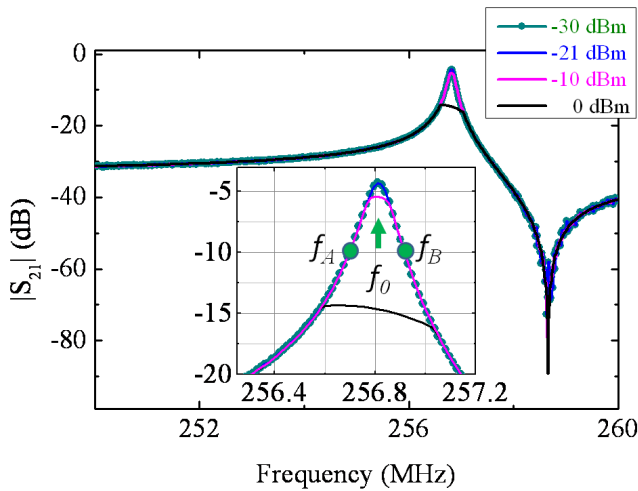}
		\caption{\label{fig:peak_powerdep} Plot of the transmittance spectrum $|S_{21}(f)|$ in the 3rd harmonic frequency ($f_0 = 256.81$ MHz) of the spiral resonator for a set of rf input powers at $T$ = 4.8 K. The inset is a close-look of the transmittance spectrum near the resonance that corresponds to input power values from -30 dBm to -10 dBm. Note that the $|S_{21}(f)|$ curves overlap, until it sharply switches to a single hot-spot resistive state at -10 dBm and progressively adds more dissipation at an input power of 0 dBm. The frequencies $f_A$ and $f_B$ are used to create images of $PR_R$ and $PR_X$ in Section \ref{PR Image Analysis}.   }
\end{figure}

 Experimentally, there are a number of competing mechanisms that may easily mask the ABS response in the HTS spiral sample. The aNLME effect is weak enough at nonzero temperature and, therefore, large current densities are required to measure very small changes in $\lambda(T,j)$. This means that extrinsic sources of nonlinearity, such as the presence of grains, grain boundaries and local structural defects, may obscure the intrinsic anisotropy of YBCO, making the LSM analysis extremely challenging. Thus, it is important to identify the upper (critical) limit of driving RF power $P_{RF}$, before extrinsic nonlinear mechanisms are activated. For a rough estimation, one can find the smallest amplitude of the input RF excitation that degrades the Lorentzian shape of the resonant transmission profile in the mode under examination.\cite{Kurter2011PRB} Figure \ref{fig:peak_powerdep} illustrates the power dependent variations of $|S_{21}(f,P_{RF})|$ for the example of the third harmonic resonance at 4.8 K. A detailed view of the upper part of the profile is pictured in the inset. As expected, the resonant peaks of transmission curve $|S_{21}(f,P_{RF})|$ are almost overlapped keeping their original form of the same Lorentzian function (blue symbols in the inset) at the highest $P_{RF}$ up to -10 dBm. Those curves clearly demonstrate the stable (relative to RF current) Meissner state where YBCO remains in the hot-spot-free superconducting phase.\cite{Kurter2011PRB} At a critical input power ($P_c\sim  -10$ dBm in this case), $|S_{21}(f,P_{RF})|$ makes a sharp transition from one Lorentzian curve onto another with higher insertion loss and lower quality factor $Q$ as frequency is scanned near resonance (the magenta curve in Fig. \ref{fig:peak_powerdep}).\cite{Zhuravel2012PRB} With further increasing input power, this transition occurs at progressively lower frequencies where the dissipative mechanism is activated (the black curves in Fig. \ref{fig:peak_powerdep} for a power of 0 dBm). To guarantee characterization of the aNLME in the Meissner state of the YBCO spiral, the bulk of the LSM results was obtained at $P_{RF}=-21$ dBm, ten times smaller than the critical RF power of the sample under investigation. 

\subsection{Spatially resolved photoresponse results} \label{Spatially resolved results}
\par The method of low-temperature Laser Scanning Microscopy (LSM) has been applied to identify the intrinsic origin of the anisotropic ABS response. The sample of interest is excited at or near resonance by an applied RF or microwave signal of frequency $f_0$ (Fig. \ref{fig:LSM_setup}). While the RF currents are oscillating in a standing wave mode the sample is perturbed by a focused laser probe. The resulting localized heating causes changes in the local electrodynamic properties of the material. These changes result in a change of resonant frequency and/or quality factor of the resonant device. This in turn changes the global transmission response $S_{21}(f)$ of the device. The LSM technique images the photo-response $PR\sim P_{RF}(\partial ||S_{21}||^2/\partial T) \delta T$, where $\delta T$ is the magnitude of local temperature oscillation due to amplitude modulated laser heating.\cite{Zhuravel2013,Zhuravel2006LTP} One can choose the stimulus frequency $f_0$ to be near the points where $\partial ||S_{21}||^2/\partial T$ is maximized. The principle of the LSM is to scan the surface of the superconducting spiral under test in a raster pattern with the focused laser beam, while detecting the $PR(x,y)$ as a function of laser spot position $(x,y)$. The photo-response map is transformed into a 2D array of digital data that are stored in the memory of a computer as contrast voltage $\delta V(x,y)$ for building a 2D LSM image of RF properties of the superconductor. In our experiments, the power of the laser is fixed at $P_L=1.6$ $\mu$W and is low enough to produce minimal perturbation on the global RF properties of the YBCO spiral resonator. The intensity of the laser is TTL modulated at a frequency of $f_M=100$ kHz creating the thermal oscillation probe in the best laser beam focus. In such a way, only the ac component of the LSM PR is detected by a lock-in technique to enhance the signal-to-noise ratio and hence the contrast of the resulting images. A number of specific schemes for the LSM optics and electronics designed for the different detection modes have been published elsewhere\cite{Zhuravel2013,Zhuravel2006LTP,Zhuravel2006JSupercond,Sivakov1994,Kurter2012APL} and it is not a subject of discussion here.

\par A simplified schematic diagram of the experimental LSM setup is pictured in Fig. \ref{fig:LSM_setup}. To form a Gaussian laser probe of 10 $\mu$m diameter, the collimated beam of the diode laser (wavelength 640 nm, maximum power 50 mW) is focused on the spiral surface with an ultra-long working distance 100 mm, 2x, NA = 0.06 objective lens. Two plane mirrors in orthogonal orientation, moved by galvano scanners, are used for the probe $(x,y)$ rastering across a 5$\times$5 mm$^2$ area with the spatial accuracy of $\pm$1 $\mu$m. While scanning, the YBCO spiral is stimulated by a microwave synthesizer (Anritsu MG37022A) at one of two driving frequencies $f_A=f_0-\Delta f$ or $f_B=f_0+\Delta f$ which are symmetrically positioned by $\Delta f$ below (at $f_A$) or above (at $f_B$) the frequency $f_0$ of the studied resonance (see inset in Fig. \ref{fig:peak_powerdep}). Here, $\Delta f$ is a half width at half maximum (HWHM) of the $S_{21}(f)$ spectral curve near the resonance frequency $f_0$. A crystal diode detects the RF amplified changes in laser-modulated RF transmitted power at those $f_A$ or $f_B$ frequencies and creates an output voltage $V$. These images of the LSM PR are then processed into separate resistive $PR_R(x,y)$ and inductive $PR_X(x,y)$ components, which will be discussed in detail at Sec. \ref{PR Image Analysis}. 

\par There are two complementary LSM modes, which were used for the presentation of experimental data. The first (2D imaging) mode allows spatially resolved visualization of modulation in the surface ABS response due to the illumination of the laser probe as a function of probe position $(x,y)$ on the sample area. Assuming we have information of the boundary surface which host ABS, the resulting LSM images in this situation give information about the in-plane anisotropy of the gap structure. The second (local probing) mode enables one to get the RF power ($P_{RF}$) and/or temperature dependence of the ABS response at any fixed position of the probe on the sample surface including both nodal and anti-nodal lines (e.g. points 1 and 2 in Fig. \ref{fig:sprial_shape}). Therefore, the 2D imaging mode was used to establish the locations of detailed probing experiments in precisely defined positions of interest. 

\begin{figure}
		\includegraphics[width=0.8\columnwidth]{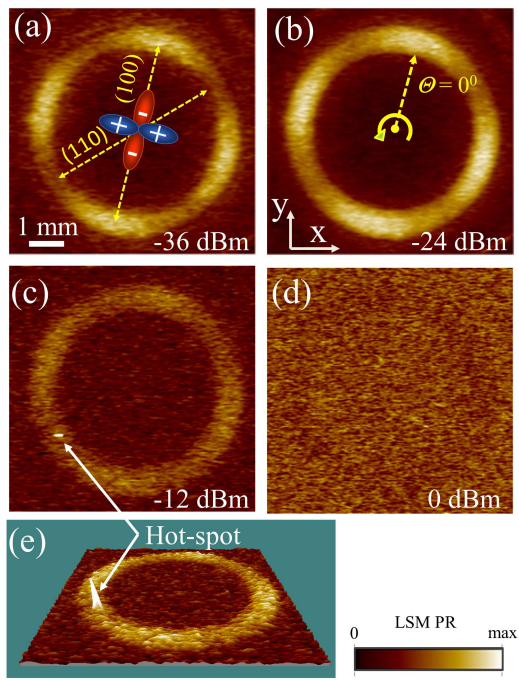}
		\caption{\label{fig:2Dimages} 2D LSM PR images of the YBCO/MgO spiral resonator at $T=4.8$ K for input $P_{RF}$ of (a) -36 dBm, (b) -24 dBm, (c) -12 dBm, and (d) 0 dBm; (e) 3D LSM PR image showing a hot spot. Note that color scheme for each plot is determined by minimum and maximum value of PR at each plot. Dashed arrows in (a) show directions of (110) and (100) crystallographic planes of YBCO where the directions of the current at those locations of the spiral are parallel to gap nodal and anti-nodal direction. In addition, the zero position of $\Theta$  and its direction of rotation are shown in (b).}
\end{figure}

\par Figure \ref{fig:2Dimages} shows RF power dependent modification of 2D LSM PR images acquired in the area of the YBCO spiral at four different values of applied $P_{RF}$ in the range from -36 dBm to -6 dBm at $T=4.8$ K (which is well below $T_m = 7.3$ K). The images were recorded at a frequency $f_A$ at a point in $|S_{21}(f)|^2$ that is 3 dB below the peak of the third resonance mode ($f_0$ = 256.8 MHz, see Fig. \ref{fig:peak_powerdep}). Brighter regions in the images correspond to those areas of the spiral yielding a higher laser probe induced $PR(x,y)$. The first measurable $PR(x,y)$ appears at $P_{RF}=-36$ dBm (see Fig. \ref{fig:2Dimages}(a)) as an anisotropic pattern of LSM photoresponse demonstrating a 4-fold angular ($\Theta$) symmetry. As one might expect, there is a strong general correlation between the $PR(\Theta)$ distribution and angular position of the gap nodal (110) and antinodal (100) planes of the $c$-axis oriented YBCO film.\cite{Zhuravel2006LTP} This is clearly illustrated in Fig. \ref{fig:2Dimages}(a) through the linking of the LSM image with the $ab$ crystallography of YBCO as marked by arrowed dashed lines along with $d_{x^2-y^2}$ gap orientation at the figure center. Here, the $a$, $b$ axis directions of the YBCO film are determined from the directions of the $a$, $b$ axis of the substrate assuming they are parallel to the crystallographic axis of the film, and also from the direction of twin boundaries which are supposed to be aligned with the (110) direction. Once the $a$, $b$ axis directions are determined, one can determine the directions of $k_x$ and $k_y$ in momentum space in the images and hence can determine the gap nodal direction ($k_{xy}$) and antinodal direction ($k_x, k_y$). Note that in the spiral sample, the direction of the current is tangential to the spiral line. Therefore, the relative direction between the local current density to the gap node at a certain position on the spiral can be easily determined.  

\par In the next example, Fig. \ref{fig:2Dimages}(b) shows the pattern of $PR(x,y)$ at input power of -24 dBm demonstrating an unchanged form of the spatially modulated response for undercritical excitation. This anisotropic NLME pattern keeps the same spatially aligned form up to $P_{RF}$ = -12 dBm (63 $\mu$W) when the first detectable distortion of the LSM image is visible through the effect of the nonsuperconducting ``hot spot'' formation. The hot spot arises at spatially localized weak links and microscopic defects in several areas of YBCO having different microwave properties from the rest of the film (see Fig. \ref{fig:2Dimages}(c) and 3D image of pointed area by the arrow in Fig. \ref{fig:2Dimages}(e)).\cite{Zhuravel2010JAP} At even higher RF powers, multiple dissipative hot-spot domains are activated, eventually leading to degradation of the resonant response and disappearance of LSM PR amplitude as seen in Fig. \ref{fig:2Dimages}(d). 

\begin{figure}
		\includegraphics[width=1\columnwidth]{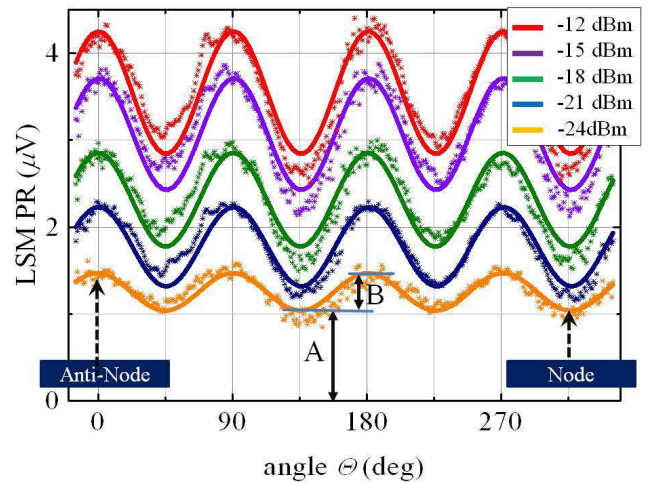}
		\caption{\label{fig:expPRangdep} Plot of radially averaged and unwrapped LSM PR (symbols) vs. angle for a series of RF powers exciting the YBCO/MgO spiral at 4.8 K, along with corresponding fits (solid lines) to the simple $d_{x^2-y^2}$ model of angular dependent PR. }
\end{figure}

\par Close examination of Fig. \ref{fig:2Dimages} shows that there are two interesting observations to be made. First, at low field RF excitation of the YBCO spiral, the angular position of the peak amplitudes of LSM $PR(\Theta)$ are aligned along the antinodal ((100),(010)) lines, which will be explained in detail in Sec. \ref{Theory}. The other interesting observation is that LSM PR images at $T<T_m$ become blurred (see Fig. \ref{fig:transmission_profile}(e)) in comparison with a sharp view of the standing wave pattern which has been obtained for the same resonance mode at $T>T_m$ (see Fig. \ref{fig:transmission_profile}(d)). This feature is mainly due to an increased thermal healing length of the laser probe due to increase in thermal boundary resistance between the film and substrate at low temperature, which in turn decreases the spatial resolution of the probe.\cite{Zhuravel2006LTP,Zhuravel2006APL}

\par Figure \ref{fig:expPRangdep} shows the angular ($\Theta$) dependence of the radially ($\rho$) averaged $PR$ for a series of fixed values of $P_{RF}$. Experimental data of $PR(\Theta)$ for a YBCO/MgO thin film spiral resonator were extracted from a set of 2D images taken in the 3rd harmonic mode at 256.8 MHz (see Fig. \ref{fig:2Dimages}). Both the zero-angle position and angular direction for $PR$ unwrapping are shown in Fig. \ref{fig:2Dimages}(b). Locations of the closest (to $\Theta=0^\circ$) nodal and anti-nodal lines are marked in Fig. \ref{fig:expPRangdep} by dashed arrow lines. For clarity, results for each specific $P_{RF}$ are symbolized by individual colors as shown in the legend. The same colors specify the solid line fitting curves that present $PR(\Theta)$ in the frame of a simple model of $PR(\Theta)=A+B\sin^2(2\Theta)$ which gives a very good fit to the angular dependence data. Here, A is the offset and B is the amplitude of anisotropy of $PR(\Theta)$ as shown in Fig. \ref{fig:expPRangdep}. As applied $P_{RF}$ increases, so do the fit values of A and B, which means both of them are power dependent. Nonetheless, the same angular modulation of the LSM $PR \sim\sin^2(2\Theta)$ remains evident independent of $P_{RF}$, completely determining the general description at any RF power level. Physically, the two extreme locations of $PR(\Theta)$ on the surface of the YBCO spiral are most interesting. The local probing LSM measurements were carried out with the object of detailed analysis on those features of YBCO spiral PR anisotropy with respect to the amplitude of the microwave field.

\subsection{RF power dependence of photoresponse}
\begin{figure}
		\includegraphics[width=0.8\columnwidth]{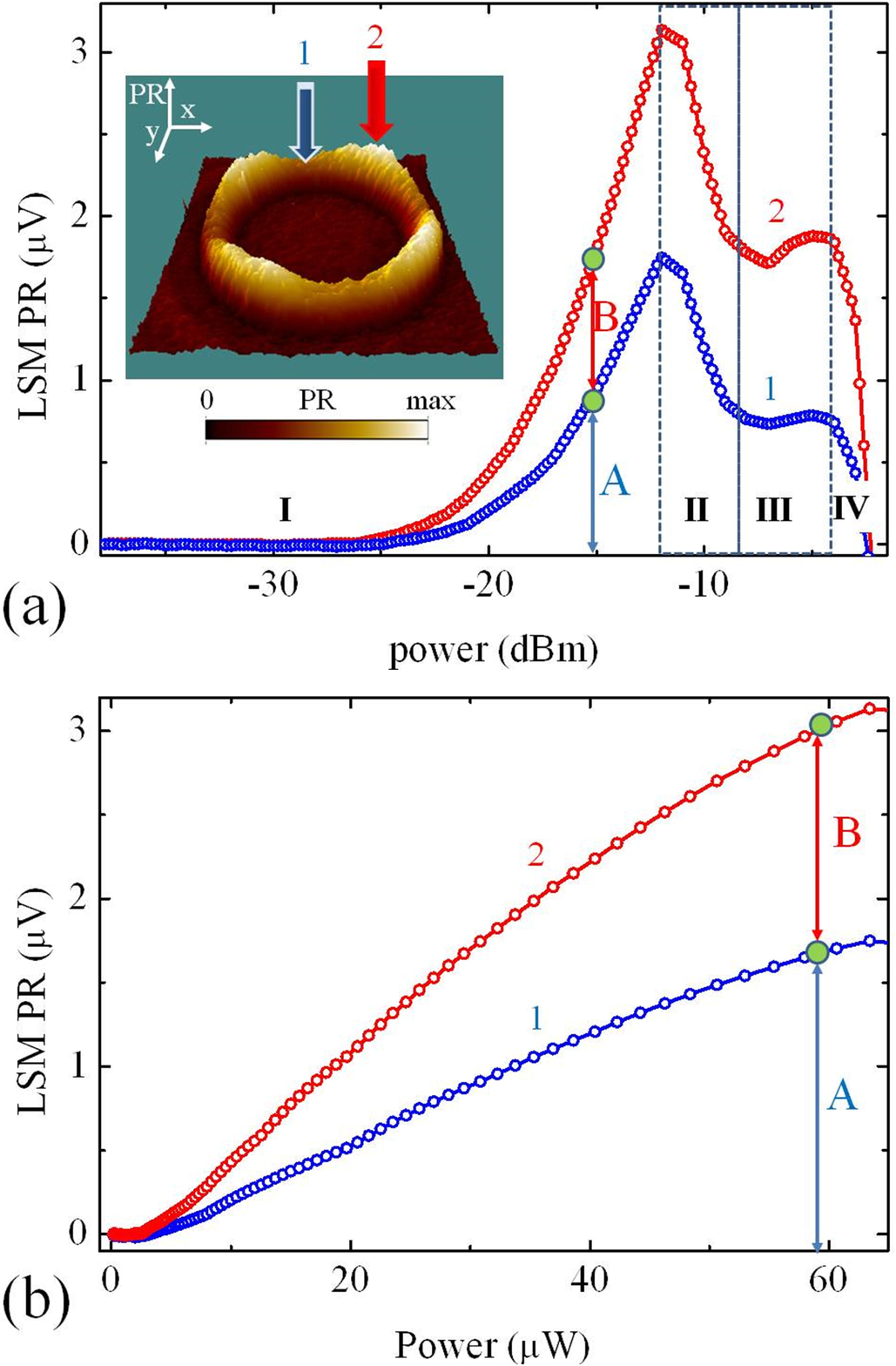}
		\caption{\label{fig:expPRqdep} Plot of LSM PR vs. $P_{RF}$ on a logarithmic scale, taken in nodal (curve 1) and antinodal (curve 2) positions of YBCO/MgO spiral resonator at $T = 4.8$ K and at $f_0=256.8$ MHz. Here, A is PR measured at the nodal position, and B is the difference in PR between antinodal and nodal positions. Positions of the laser probe that have been used to record the data are marked by arrows in the inset 3D image showing LSM PR visualized at $P_{RF}=-21$ dBm at $T = 4.8$ K; (b) Detailed view of the same plot on a linear scale of $P_{RF}$, measured in the low field Meissner region I of RF excitation. }
\end{figure}
\par Curve 1 (Blue) in Fig. \ref{fig:expPRqdep}(a) shows the RF power dependence of the LSM PR which is measured at a fixed position of the laser probe that is focused at point 1 (see Fig. \ref{fig:sprial_shape}). The position of point 1 coincides with gap nodal line (110) of YBCO in-plane crystallography. The location of the probe is shown by the blue arrow 1 in the inset of Fig. \ref{fig:expPRqdep}(a) that presents a 3D LSM PR image which is acquired at $P_{RF} = -21$ dBm at $T = 4.8$ K. The local probing was done in the 3rd harmonic mode at 256.8 MHz at $T=4.8$ K. Experimental data of the LSM PR vs. $P_{RF}$ were recorded by a stepwise changing of the input RF power with equal steps of 0.1 dBm. By refocusing the laser probe to point 2 (See Fig. \ref{fig:sprial_shape}), $PR(P_{RF})$ data were obtained in the same way at the location of an antinodal line (see red curve 2 in Fig. \ref{fig:expPRqdep}(a)). As expected from the 2D images (see Figs. \ref{fig:2Dimages} and \ref{fig:expPRangdep}), both A and B are a monotonically increasing functions of $P_{RF}$ at low magnitude of RF fields in region I. The same plot looks more informative on a linear $P_{RF}$ scale as shown in Fig. \ref{fig:expPRqdep}(b). Here, the angularly localized components of gap nodal (red curve 1) and anti-nodal (blue curve 2) contributions to $PR(P_{RF})$ are plotted solely in region I, restricting the power scale to a maximum value of $P_{RF}\sim 63$ $\mu$W ($=-12$ dBm) that corresponds to initialization of the first hot-spot nucleation.\cite{Kurter2011PRB} As the power increases, the number of hot spots increases, producing a nonlinear increase of the surface resistance $R_S(j_{RF})$ that, in turn, causes degradation of the $Q$-factor in $|S_{21}(f,P_{RF})|$ which decreases the $PR(P_{RF})$ magnitude (see region II in Fig. \ref{fig:expPRqdep}(a)). Further increase in $P_{RF}$ (as seen in region III) causes a metamorphosis of a spatially distributed resistive structure of hot-spots into a stable pattern of normal domains that thereafter are generating an unstable overheating effect with increased power in region IV.\cite{Zhuravel2012PRB} Hence only region I is experimentally compatible with the requirement of searching for intrinsic components of an anisotropic quasiparticle (ABS NLME) and superfluid (bulk NLME) responses in this sample. Moreover, we found that the LSM probed upper limit of $P_{RF}=-12$ dBm in this case is almost two times below the critical power $P_c$ that was determined by global measurement (see above text on Fig. \ref{fig:peak_powerdep}) employing $|S_{21}(f,P_{RF})|$ analysis. This confirms once again that the LSM technique is more sensitive than global characterization, making it possible to specify experimental regions of clear observable effects with the highest precision. With this result, the previously adopted choice of $P_{RF}=-21$ dBm at a temperature of 4.8 K is adequate to study the ABS response of the YBCO spiral resonator. 

\subsection{Photoresponse image analysis} \label{PR Image Analysis}
\par
\begin{figure}
		\includegraphics[width=0.8\columnwidth]{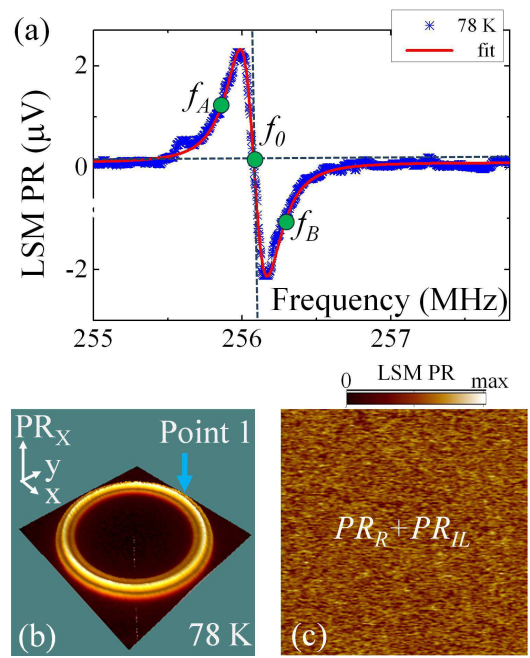}
		\caption{\label{fig:PR_78K} (a) Experimental (blue symbols) and fitting (red solid line) curves of the frequency dependent total LSM PR in Point 1 of the YBCO/MgO spiral resonator at T = 78 K and $P_{RF} = -21$ dBm. The data was obtained in nodal regions. LSM images of (b) inductive $PR_X(x,y)$ and (c) resistive $PR_R(x,y)$ components. }
\end{figure}

\begin{figure}
		\includegraphics[width=0.9\columnwidth]{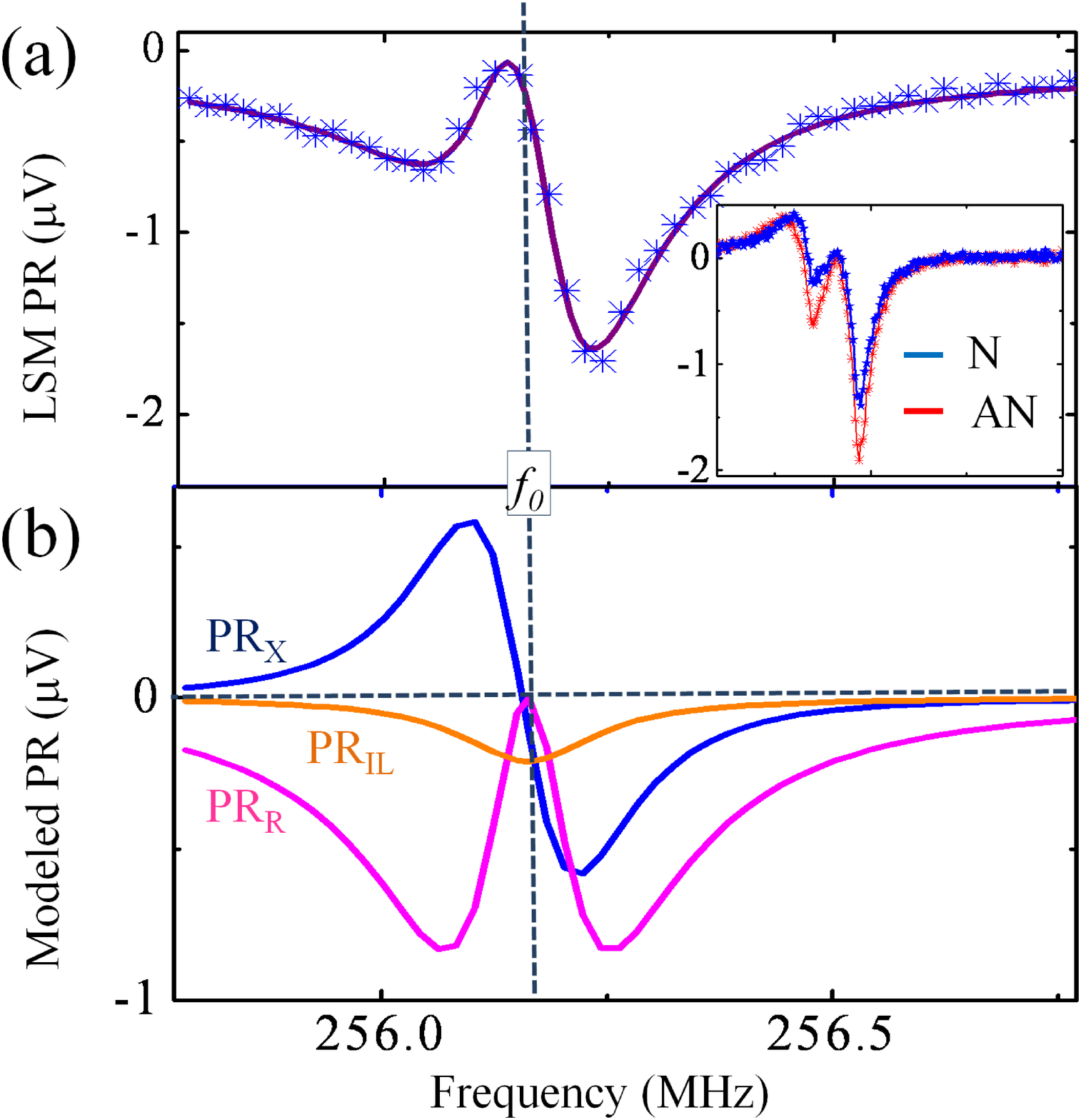}
		\caption{\label{fig:PR_4K} (a) Frequency dependence of experimental (symbols) and fitting (solid line) data of local PR of YBCO/MgO superconducting spiral sample probed at point 1 corresponding to the direction of the nodal lines. Experimental data were obtained at T = 4.8 K in the third harmonic mode at $P_{RF} = -21$ dBm; (b) result of modeling decomposition of the $PR(f)$ on individual inductive $PR_X$, resistive $PR_R$ and insertion losses $PR_{IL}$ components. Inset shows experimental plots of $PR(f)$ at nodal (N) and anti-nodal (AN) points.}
\end{figure}

 Now that the overall picture of the power dependence of PR$(\Theta)$ anisotropy has been established, a microscopic understanding of its local sources must be developed. At a fixed laser perturbation location, the LSM PR is proportional to the probe-induced changes in resonator transmittance $\delta\Vert S_{21}(f)\Vert^2$ that can be decomposed into three parts in terms of their origins. One is inductive $PR_X \propto (\partial f_0/\partial T)\delta T$, another is resistive $PR_R \propto \partial(1/2Q)\delta T$, and the other is insertion loss $PR_{IL} \propto (\partial \bar{S}^2_{21}/\partial T)\delta T$ responses. Here, $\delta T \sim 10$ mK is the local temperature oscillation amplitude underneath the laser probe and $\bar{S}_{21}$ is the maximum of the transmission coefficient as a function of frequency. Note that both $PR_R(x,y)$ and $PR_{IL}(x,y)$ are linked with several dissipation mechanisms, for example, Ohmic dissipation from quasiparticle flow $\propto \delta(j_{RF}^2(x,y)R_s(x,y))$. The $PR_X$ term is directly related to the bolometric change of energy from the kinetic inductance $E_K \propto L_K j^2_s$ of the superconducting resonator. Here, an important question arises : How much relative contribution does each PR component make in each temperature regime? By focusing the laser probe at point 1 (see Fig. \ref{fig:sprial_shape}) on the nodal direction, we extracted the local values of these significant components of LSM PR at two different temperatures characterizing response of the YBCO spiral resonator in (i) the isotropic Meissner effect regime at T = 78 K (see Fig. \ref{fig:PR_78K}(a)) and (ii) the anisotropic NLME regime at T = 4.8 K (see Fig. \ref{fig:PR_4K}(a)). Note that this temperature dependent isotropy/anisotropy of the NLME originates from that of the nonlinear Meissner coefficient.\cite{Dahm1996,Zhuravel2013} Both experiments were carried out at the same $P_{RF}$ = -21 dBm ($\ll P_c$) in the 3rd harmonic mode of the spiral resonance.

\par The frequency dependence of the total LSM PR at 78 K is symbolized by the blue stars in Fig. \ref{fig:PR_78K}(a). As expected, at reduced temperature $T/T_c > 0.5$, the $PR(f)$ can be approximated by fitting (red solid line) to only a $PR_X(f)$ component. It is apparent that precisely the same profile of the local photo-response has also been measured at anti-nodal point 2 (See Fig. \ref{fig:sprial_shape}) and, thus, it is not presented here. In addition, three LSM images of $PR(x,y)$ were obtained at frequencies $f_A$, $f_B$, and $f_0$ at the same experimental conditions to extract the 2D spatial distribution of the individual components of PR using the procedure of spatially-resolved complex impedance partition.\cite{Zhuravel2006LTP,Zhuravel2006JSupercond,Zhuravel2006APL,Zhuravel2007MSMW,Zhuravel2010MSMW,Zhuravel2007IEEE} As is evident from the restored LSM image in Fig. \ref{fig:PR_78K}(c), the dissipative response $PR_R$ (+$PR_{IL}$) introduces no contribution, hence the total PR is dominated by $PR_X(x,y)$ in the linear Meissner state at 78 K. Another important observation can be shown from Fig. \ref{fig:PR_78K}(b) where the inductive component,\cite{Culbertson1998,Zhuravel2002APL}
\begin{equation}
PR_X(\rho,\Theta) \propto \lambda^2(\rho,\Theta)j^2_s(\rho,\Theta)\delta\lambda(\rho,\Theta)
\end{equation}
looks almost isotropic, demonstrating a clear pattern of superfluid distribution in an undistorted standing wave. This means that in the linear RF regime, (i) $PR_X$ is independent of in-plane direction of the $j_s$ even as the superfluid flows along/across the Cu–O bonds and simultaneously (ii) so is $\lambda$.

\par The blue symbols in Fig. \ref{fig:PR_4K}(a) show experimental data of PR vs. frequency $f$ for a YBCO spiral sample with anisotropic response at T = 4.8 K. This result is derived from a local probing at a nodal line position ($\Delta=0$) at point 1. The general shape of the curve becomes complex for $T<T_m$ and, in addition to that, the shape changes when the same measurement is repeated at the position of the anti-nodal (AN) lines ($\Delta$ = max). To understand these features, we decomposed the nodal LSM PR to its separate components as indicated in Fig. \ref{fig:PR_4K}(b). The sum of the fractional components over all of inductive (blue), resistive (magenta) and insertion loss (light brown) response is presented in Fig. \ref{fig:PR_4K}(a) as the fitting (red line) curve. Note that the dissipative $PR_R$ component is large ($PR_R/PR_X\sim 1.4$), contrary to the basic RF properties of superconductors in the Meissner state which produces dominant inductive $PR_X$ response at $P_{RF}\ll P_c$. Moreover, this $PR_R$ component still persists (with ratio of $PR_R/PR_X\sim 1.2$) even in the case of AN response (see inset in the Fig. \ref{fig:PR_4K}(a)) despite its current flow in the direction of a fully open superconducting gap. A possible source of this effect is the strong concentration (localized within the coherence length $\xi$) of paramagnetic normal fluid current at the (110) surfaces of YBCO. This, in turn, produces a substantial increase of resistive loss proportional to the normal current squared showing indirect evidence for the nonlinear paramagnetic response from the (110) boundary surface.

\begin{figure}
		\includegraphics[width=1\columnwidth]{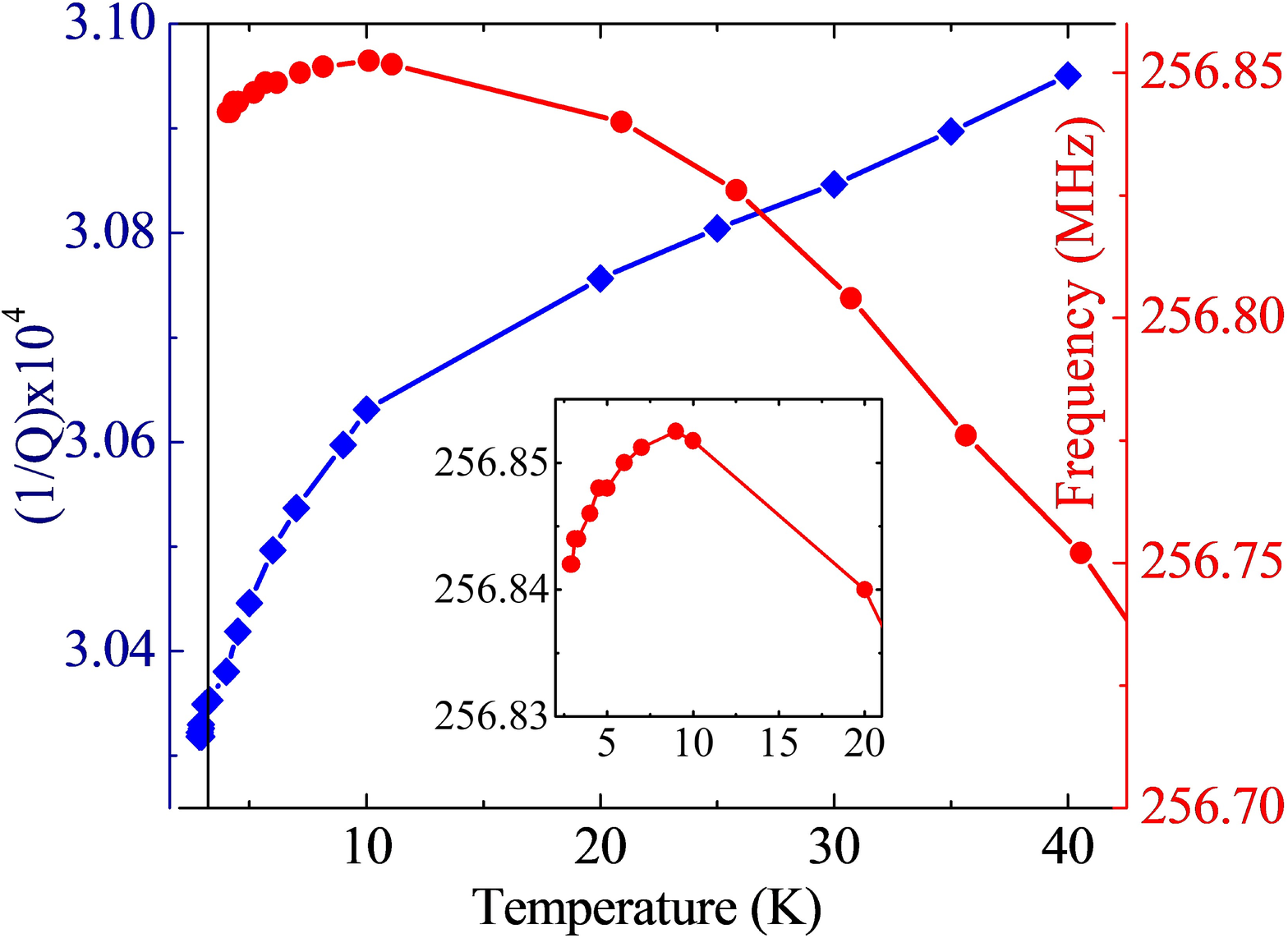}
		\caption{\label{fig:f0andQ} Plots of resonance frequency (red circles) and inverse quality factor (blue diamonds) vs. temperature for incident power of -21 dBm of YBCO/MgO superconducting spiral sample. Detailed view of low-temperature data for $f_0(T)$ is shown in the inset.}
\end{figure}

\subsection{ABS contribution to the penetration depth}
\par For an ABS to exist at the boundary surface such as a twin domain boundary, a quasiparticle should experience a $\pi$ phase difference of the order parameter before and after the reflection at the boundary surface. The twin boundary in a YBCO film is oriented in the (110) direction which means at any incident angle, the quasiparticle experiences such a phase difference. Therefore, the prerequisite for the formation of the ABS is always fulfilled. In most cases, a YBCO thin film has twin boundary separation less than 100 nm.\cite{Streiffer1991APL} Hence there is no need to control the in-plane direction of the applied field to see ABS paramagnetic response in the experiments with the superconducting YBCO spiral because of the abundance of twins. Moreover, the response is multiplied several tens of times due to the repetition of the fourfold gap configuration within all turns of the spiral. Thus, one can expect a significant ABS response from a YBCO thin film spiral sample. In this case, a low-temperature upturn of magnetic penetration depth would be reasonable evidence of strong paramagnetic Meissner effect from ABS.\cite{Walter1998PRL,Carrington2001,Prozorov2006SUST}

\par In Fig. \ref{fig:f0andQ}, the 3rd harmonic resonance frequency of the YBCO/MgO spiral resonator is depicted as a function of temperature. This is the global response of the resonator in the absence of laser perturbation. This frequency increases at fixed $P_{RF}=-21$ dBm $\ll P_c$ as T decreases down to 10 K demonstrating the expected linear-response changes of inductance and effective magnetic penetration depth $\lambda_{eff}(T)$ at $T<T_m$. The resonant frequency in this case can be described well by the usual theoretical temperature dependence $f(T)=f(0)\left[1+2\lambda(T)\coth(t/\lambda(T))/d\right]^{-1/2}$,\cite{Anlage1989APL,Talanov2000RSI} where $d$ is a characteristic length scale of the resonator, $t$ is the thickness of the YBCO film, $f(0)$ is the resonant frequency of a perfectly conducting ($\lambda=0$) material, and magnetic penetration depth is approximated by $\lambda(T)\simeq \lambda_0[1-(T/T_c)^2]^{-1/2}$. However, a maximum of $f_0(T)$ is observed for lower $T$ between 10 K and 5 K, and frequency shift reverses for $T<5$ K as the temperature further decreases. This non-monotonic temperature dependence can be attributed to one or more of five mechanisms. First, the low temperature upturn of the screening length due to impurity paramagnetism,\cite{Mercaldo2000,Prozorov2006SST} second due to the paramagnetic properties of the ABS that form and become stronger at low temperatures,\cite{Carrington2001,Walter1998PRL,Barash2000,Zare2010PRL} third due to the temperature dependent NLME,\cite{Yip1992,DXu1995,Dahm1996,Dahm1997,Dahm1999} fourth due to dielectric microwave losses in the substrate,\cite{Hein2002APL,O’Connell2008APL,ZhouPinJia2014CPL} and fifth due to increase of dissipative losses with decreasing of temperature. In the following theory section, it will be shown that the theoretical estimate for photoresponse, attributing its origin to the ABS response, makes a very good agreement with the experimental data at the temperature regime where the reverse shift of $f_0(T)$ happens, which supports the scenario of the reverse shift arising from ABS response.

\section{Theory}\label{Theory}
\par
\begin{figure}
		\includegraphics[width=0.9\columnwidth]{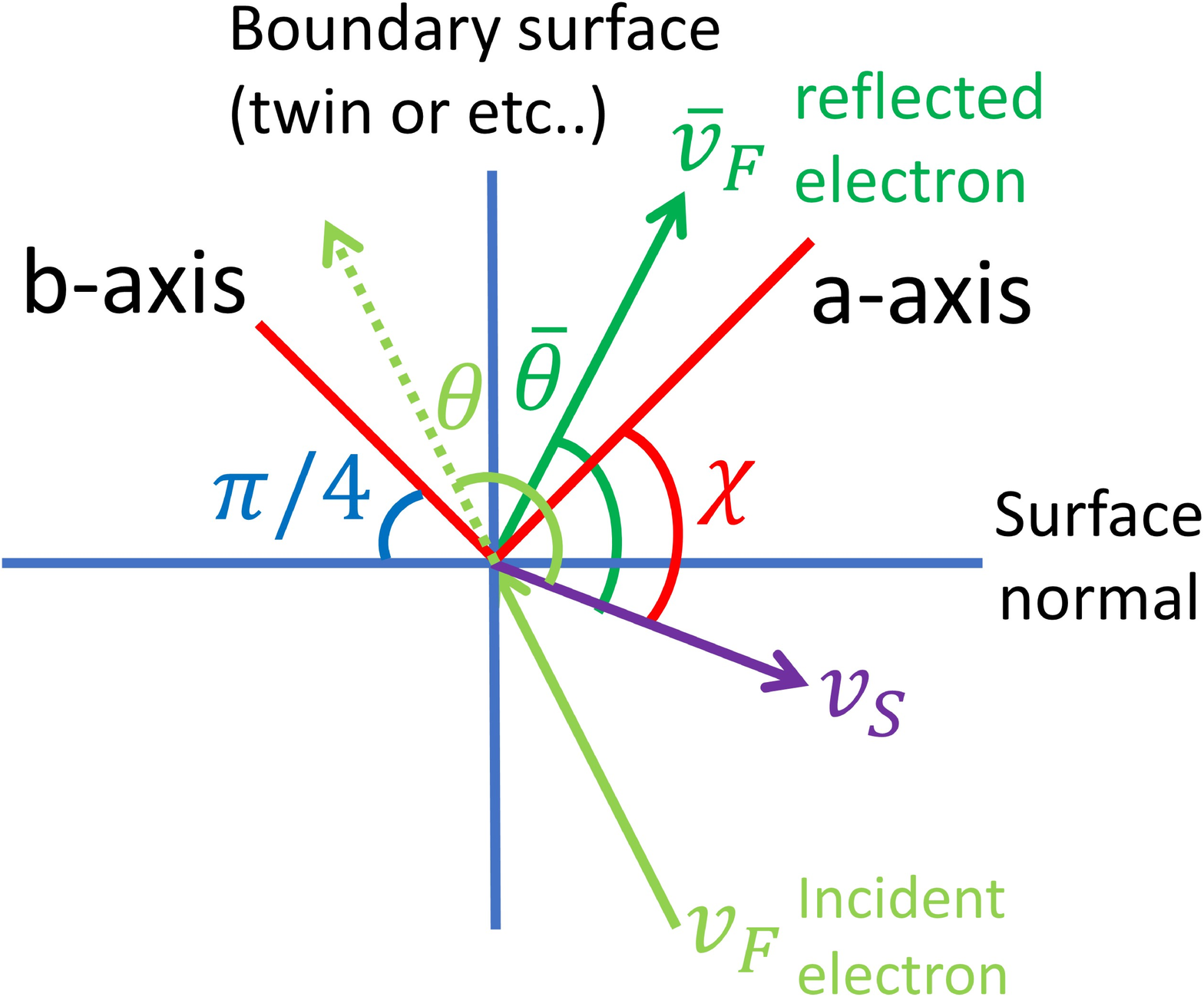}
		\caption{\label{fig:geometry_setup} Diagram showing the geometry setup of the sample system. The vertical blue line is the boundary surface, which is a twin boundary in the YBCO sprial sample. The red lines are the a and b-axis directions of the sample, which make a $\pi/4$ angle to the boundary surface. The green arrows show the direction of an incident ($\vec{v}_F$) and reflected ($\vec{\bar{v}}_F$) quasi particle from the Andreev bound state at the surface. The purple arrow is the direction of superfluid $\vec{v}_s$ driven by the external RF field. $\theta$ (or $\bar{\theta}$) is the angle between $\vec{v}_F$ (or $\vec{\bar{v}}_F$) and $\vec{v}_s$ (see green arcs). $\chi$ is the angle between the a-axis direction and $\vec{v}_s$. Since $\vec{v}_F$ and $\vec{\bar{v}}_F$ are mirror images of each other through the boundary surface, $[(\theta-\chi)+(\bar{\theta}-\chi)]/2=\pi/4$. Note that as one moves around the spiral, the direction of $\vec{v}_s$ changes but the direction of the twin surface and a,b-axis directions of the sample do not change.  }
\end{figure}

 In this section, a microscopic model is introduced to describe how a $d_{x^2-y^2}$ superconductor sample with a twin boundary, which can host Andreev bound states (ABS), responds to external RF magnetic field. Then, from the RF field response of the sample, the anisotropy (angular dependence) and input RF power dependence of the photoresponse will be estimated and the results will be compared to experimental data. First, when an external RF magnetic field is applied to such a sample, it induces current both in the bulk and on the boundary surfaces of the sample. The transport phenomena in a superconductor can be described by a quasi-classical Green function in Nambu space $\hat{G}(r,\hat{\vec{v}}_F,\omega)=\left( \begin{smallmatrix} g & f \\
f^\dagger & g^\dagger \end{smallmatrix} \right)$ which satisfies the Eilenberger equation.\cite{Eilenberger1968,Kulik1978,Belzig1999,Agassi2012PhysicaC} Here, $g$ and $f$ are normal and anomalous components of the Green function. The induced current under the external magnetic field can be calculated from this Green function\cite{Kolesnichenko2004,Kolesnichenko2003PRB,Shevchenko2009,Shevchenko2006PRB}. The resulting current density is given by
\begin{equation} \label{current_density}
j(r)=-j_0\frac{T}{T_c}\sum_{\widetilde{\omega} >0} \langle \hat{v}_FIm  g(r,\hat{v}_F,\widetilde{\omega})\rangle_{v_F},
\end{equation}
where $j_0=4\pi eN(E_F)v_FT_c$ and $N(E_F)$ is the density of states at the Fermi energy, $r$ is the distance from the boundary surface, $\langle ... \rangle_{v_F}$ represents averaging over the Fermi surface, $\hat{v}_F=\vec{v}_F/v_F$ is the unit vector along the direction of the Fermi velocity, and $\widetilde{\omega}=\omega_n+i\vec{p}_F\cdot\vec{v}_s$ represents the Matsubara frequencies under the external magnetic field where $\vec{v}_s$ is superfluid velocity and $\omega_n=\pi T(2n+1)$. In the case when the boundary surface is aligned with the (110) crystallographic direction, which is true for a twin boundary in YBCO, the normal component of the Green function at the surface $g(0)$ and the homogeneous bulk $g(\infty)$ are obtained as
\begin{gather}
g(0)=\frac{\widetilde{\omega}(\Omega+\overline{\Omega})}{\Omega\overline{\Omega}+\widetilde{\omega}^2+\Delta\overline{\Delta}}, \\
g(\infty)=\frac{\widetilde{\omega}}{\Omega}.
\end{gather}
Here, $\Delta=\Delta_0(T,\vec{v}_s)\cos 2(\theta-\chi)$ is the angle dependent order parameter where $\Delta_0(T,\vec{v}_s)$ is the magnitude of the order parameter of a bulk $d_{x^2-y^2}$ superconductor at temperature $T$ and superfluid velocity $\vec{v}_s$, which can be obtained by solving the self-consistent gap equation. Here, as seen in Fig. \ref{fig:geometry_setup}, $\theta$ is the angle between $\vec{v}_F$ and the superfluid velocity $\vec{v}_s$, and $\chi$ is the angle between $\vec{v}_s$ and the a-axis direction of the YBCO film (or gap antinode direction equivalently), which will be mapped into position angle $\Theta$ in the spiral (Fig. \ref{fig:sprial_shape}). $\Omega=\sqrt{\widetilde{\omega}^2+\Delta^2}$ is the quasi-particle energy spectrum. Note that barred quantities represent those after reflection from the surface boundary and unbarred quantities represent those before reflection, which means $(\theta-\chi)+(\bar{\theta}-\chi)=\pi/2$. Therefore,
\begin{equation}\label{orderparam_and_excitation_spectrum}
\begin{split}
\overline{\Delta}&=\Delta_0(T,\vec{v}_s)\cos 2(\pi/2-(\theta-\chi)) \\
&=\Delta_0(T,\vec{v}_s)\cos (\pi-2(\theta-\chi))=-\Delta, \\
\overline{\Omega}&=\sqrt{\widetilde{\omega}^2+\overline{\Delta}^2}=\sqrt{\widetilde{\omega}^2+(-\Delta)^2}=\Omega. 
\end{split}
\end{equation}

\par
\begin{figure}
		\includegraphics[width=1\columnwidth]{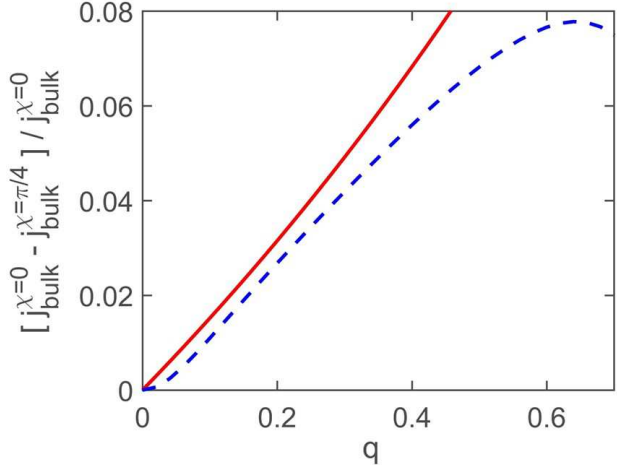}
		\caption{\label{fig:j_bulk_aniso} Anisotropy ratio in the bulk Meissner current density, written as the relative value of $j_{bulk}$ for the angles $\chi=0$ and $\pi/4$, as a function of superfluid momentum $q=p_F v_s/\Delta_0$. The solid line illustrates Eq. (\ref{aniso_ratio_jbulk}) which ignores superfluid momentum dependence of the order parameter $\Delta_0=\Delta(T,\vec{v}_s=0)$, while the dashed line is the result of the numerical calculations, which take into account the dependence of $\Delta_0=\Delta(T,\vec{v}_s)$, demonstrated at low temperature, $T/T_c=0.05$. }
\end{figure}

 With the Green function presented above, the current density of the bulk Meissner state $j_{bulk}$ and of the surface Andreev bound state $j_{surf}$ at various experimental parameters can be calculated. For a validation of the presented numerical scheme, its result is compared to the famous Yip and Saul's result\cite{Yip1992} where they derive a theoretical formula for the superfluid momentum $q(=p_Fv_s/\Delta_0)$ dependence of the anisotropy ratio of $j_{bulk}$, defined as the relative value of the $j_{bulk}$ for the angles $\chi=0$ and $\pi/4$. It is given as,
\begin{equation} \label{aniso_ratio_jbulk}
\frac{j_{bulk}^{\chi=0}-j_{bulk}^{\chi=\pi/4}}{j_{bulk}^{\chi=0}}=q\frac{\sqrt{2}-1}{2\sqrt{2}-q}.
\end{equation}
 This is demonstrated in Fig. \ref{fig:j_bulk_aniso} by the solid line. In spite of the seemingly large value of this ``$\sqrt{2}$-anisotropy", Eq. (\ref{aniso_ratio_jbulk}) describes only a few-percent change for the relevant values of $q$. Note that the respective formulas in Ref. \cite{Yip1992} are obtained in the first approximation on this parameter $q$. The result from this theoretical formula Eq. (\ref{aniso_ratio_jbulk}) and the result from our numerical calculation is similar for small $q<0.3$ but starts to deviate from each other for large $q$ because the result of the numerical calculation takes into account the superfluid momentum dependence of the order parameter. Considering the $q$ dependence of the order parameter, even for higher values of $q$, the anisotropy ratio of $j_{bulk}$ does not exceed a ten-percent limit.\cite{Kolesnichenko2004,Ferrer1999,EJNicol2006PRB}

\par
\begin{figure}
		\includegraphics[width=0.9\columnwidth]{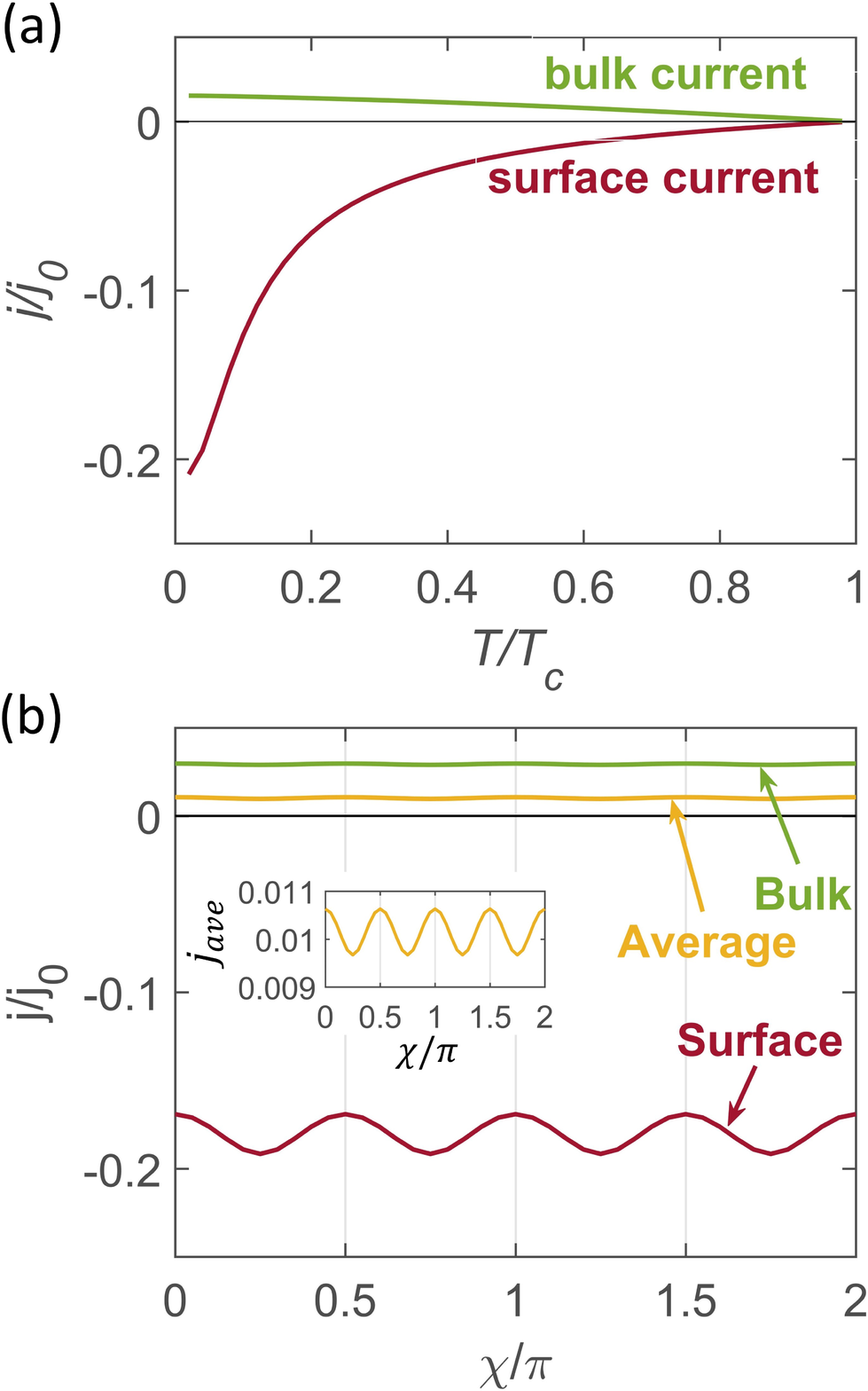}
		\caption{\label{fig:current_densities} (a) Temperature dependence of the current densities at the surface and the bulk when $v_s\parallel$ gap node ($\chi=\pi/4$) and superfluid momentum $q=p_Fv_s/\Delta_0=0.1$. The sign of the surface current is the opposite to that of the bulk diamagnetic current which implies the surface current is paramagnetic. (b) The angular dependence of the current density at the surface, bulk, and their average when $q=0.2$ and $T/T_c=0.05$. Inset is a close-up plot of average current density vs. $\chi$.}
\end{figure}

 With this validation of our calculation, the temperature-($T$) and angular-($\chi$) dependence of $j_{surf}$ and $j_{bulk}$ is presented in Fig. \ref{fig:current_densities}. As shown in Fig. \ref{fig:current_densities}(a), both of the current components increase in magnitude as temperature decreases, but the slope of increase for the case of the current at the surface is much steeper than that of the bulk current, which implies that the surface response will play a much more important role in photoresponse at low temperature. Also, note that the sign of the surface current and bulk current is opposite, which implies that the surface current is a paramagnetic current in contrast to the bulk diamagnetic current. Also note that, as shown in Fig. \ref{fig:current_densities}(b), the anisotropy of the surface current is much larger than that of the bulk current.

\par With a proper weighting factor, the average current can be calculated. Assuming that the surface paramagnetic current flows within a depth on the order of the coherence length and the bulk diamagnetic Meissner current flows within a depth on the order of the penetration depth, and they add linearly, the average current density in the sample becomes
\begin{equation}
\begin{split}
j_{ave}&\cong \frac{1}{\lambda}\int^\lambda_0 dx\left(j_{surf} e^{-x/\xi_0} + j_{bulk} e^{-x/\lambda} \right)\\
&\approx  \frac{\xi_0}{\lambda} j_{surf} + 0.5 j_{bulk}.
\end{split}
\end{equation}
Hence the contribution of the surface current relative to that of the bulk current is determined by $\xi_0/\lambda$ as a weight factor. For the case of YBCO, which is a representative type-\upperRomannumeral{2} superconductor, this ratio is quite small ($\xi_0\sim 4$ nm, $\lambda_0 \sim 160$ nm, $\xi_0/\lambda_0\sim 0.025$) so the sample gives a net diamagnetic response.

\subsection{Photoresponse estimate}
\par With these results for the RF field response of the sample, a model can be introduced to estimate the anisotropy (angular dependence) and input RF power dependence of the photoresponse. In this paper, we shall assume that the photoresponse is entirely inductive in character as a first step for comparison to data. Under the perturbation given by laser illumination, the sample response to the RF field changes, and the inductive component of this photoresponse (PR) can be estimated as\citep{Culbertson1998}
\begin{equation}
 PR \sim \delta f_0/f_0\sim -\delta W/W, \label{PR estimation}
\end{equation}
where $W$ is energy stored in both magnetic fields and kinetic energy of the superfluid. Note that the changes in the field outside the superconducting sample are marginal for small local perturbations on the sample. Therefore the contribution of the outside field on the change in stored energy $\delta W$ can be ignored and we will focus on the stored energy inside the sample.\cite{Culbertson1998} Also note that the resistive component of PR is not discussed here due to the lack of a microscopic theory which explains and estimates the dependence of the loss on various experimental parameters. If the magnetic field imposed at the surface of the film is $B_0$ and the bulk penetration depth is $\lambda$, the kinetic and magnetic field energy stored inside the sample in the wide thin film case ($t$ is comparable to $\lambda$ and $st\gg \lambda^2$ ) can be calculated as\citep{Culbertson1998}\cite{Rhoderick1962}  
\begin{equation}
 W = \int_A da \frac{B^2_0 \lambda^2}{\mu_0 t},
\end{equation}
where $t\sim 300$ nm is the thickness of the sample, $s\sim 10$ $\mu$m  is the width of the film (spiral arm), $\mu_0$ is the permeability of free space, and $A$ is area of the surface of the spiral. This area integral will be ignored below since we are interested in the angular ($\chi$) and superfluid momentum ($q$, or $P_{RF}$ equivalently) dependence of the perturbation on the local stored energy, so it is sufficient to just discuss stored energy per unit area, which we denote as $w=B^2_0 \lambda^2 /\mu_0 t$.

\par However, when there is a twin domain boundary within the sample, it hosts a paramagnetic surface current ($K_{surf}=|j_{surf} \xi_0|$) at that interface and the part of the sample nearby the twin boundary experiences an enhanced magnetic field $(B_{s0}=B_0+\mu_0K_{surf})$. We introduce a paramagnetic weighting factor $p$ which reflects the portion of the sample that experiences an enhanced field $B_{s0}$. This parameter is different for each sample depending on its twin density. With this parameter introduced, the averaged magnetic field experienced by the sample, corresponding stored energy, and change in stored energy per unit area due to the external perturbation can be written as  
\begin{gather}
B_{ave}^2 = (1-p)B_0^2+pB_{s0}^2,  \\
w = B_{ave}^2\lambda^2 /\mu_0 t, \\
\delta w = \frac{2pB_{s0}\lambda^2}{t}\delta K_{surf} + \frac{2B_{ave}^2}{\mu_0 t}  \lambda \delta\lambda.   \label{change in stored energy}
\end{gather}
The first term in Eq. (\ref{change in stored energy}) shows the contribution to nonlinear response from the surface current in an Andreev bound state (ABS) and the second term shows that from bulk current due to the nonlinear Meissner effect.

\par To estimate the photoresponse, one needs to know $K_{surf}$ and $j_{bulk}$ (which in turn gives an estimation for $\lambda$). We have already derived expression for those quantities through Eqs. (\ref{current_density}-\ref{orderparam_and_excitation_spectrum}) for the sample geometry in Fig. \ref{fig:geometry_setup}. Once the surface ($K_{surf}$) and bulk ($j_{bulk}$) current densities are calculated from the Green function, one can expand them in terms of the superfluid momentum ($q=p_Fv_s/\Delta_0(0,0)$) in the regime of $q \ll T/\Delta_0 $\citep{EJNicol2006PRB}
\begin{gather}
K_{surf}(T,q) = j_0 \xi_0 \left(\alpha_{surf} q - \beta_{surf} q^3 + \cdots\right), \label{ExpanSurf}  \\ 
j_{bulk}(T,q) = j_0\left(\alpha_{bulk} q - \beta_{bulk} q^3 + \cdots\right), \label{ExpanBulk} \\
\lambda^2(T,q) = \lambda^2(T)\left(1+b_\chi (j/j_c)^2 + \cdots \right), \label{ExpanLambda}
\end{gather}
where $\beta_{surf}$ is the surface ABS nonlinear coefficient, $b_\chi=\beta_{bulk}/\alpha_{bulk}^3$ is the bulk nonlinear Meissner coefficient\citep{Dahm1996,Zare2010PRL,EJNicol2006PRB}, and $j_c$ is the critical current density at $T=0$ K. Under illumination by a modulated scanning laser beam, these quantities are modulated ($\delta K_{surf}$,$\delta \lambda$ in Eq. (\ref{change in stored energy})). The previous experimental study\cite{Zhuravel2013} on the temperature dependence of the photoresponse and the theoretical study\cite{Zare2010PRL} on the nonlinear Meissner coefficient are consistent with a model which attributes PR to the modulation in the nonlinear terms in the above expansion (Eqs. (\ref{ExpanSurf}-\ref{ExpanLambda})). This means $\delta K_{surf} \sim -\delta \beta_{surf} q^3$, $\delta (\lambda^2) \sim \lambda^2(T)\delta b_\chi(j/j_c)^2$. Then $\delta w$, which accounts for PR, becomes
\begin{gather}
\delta w \sim -\frac{2pB_{s0}\lambda^2}{t}\delta \beta_{surf} q^3 + \frac{B_{ave}^2 \lambda^2}{\mu_0 t} \delta b_\chi(j/j_c)^2.  \label{deltaW}
\end{gather}
Here, the first term represents photoresponse from paramagnetic current in surface Andreev bound states and the second term represents that from diamagnetic Meissner current in the bulk. Note that their signs are opposite so they compete with each other. Also, $\delta\beta_{surf}(T)$ which governs the temperature dependence of the surface response shows $\sim 1/T^4$ behavior and $\delta b_{\chi}(T)$ which governs that of the bulk response shows $\sim 1/T^2$ behavior.\cite{Barash2000,Zare2010PRL} Hence at low temperature the surface response dominates and at high temperature the bulk response dominates. Also note that surface ABS PR shows a larger contribution when $v_s\parallel$ gap antinode ($\chi=0$) and bulk nonlinear Meissner effect PR shows a larger contribution when $v_s\parallel$ gap node ($\chi=\pi/4$)\citep{Zhuravel2013}. Therefore as the temperature of the sample decreases, a $\pi/4$ angle rotation of the PR image can be observed as seen from Fig. \ref{fig:transmission_profile}(d)-(e) and one can define a PR crossover temperature $T_{cross}$ as the temperature where the surface response dominant antinodal PR ($\chi=0$) starts to be larger than the bulk response dominant nodal PR ($\chi=\pi/4$) below that temperature. Thus, from the angular dependence of PR, one can tell which response dominates for a given experimental condition.

\section{Comparison of Data and Theory and Discussion} \label{Comparison of Data and Theory}
\begin{figure}
		\includegraphics[width=0.8\columnwidth]{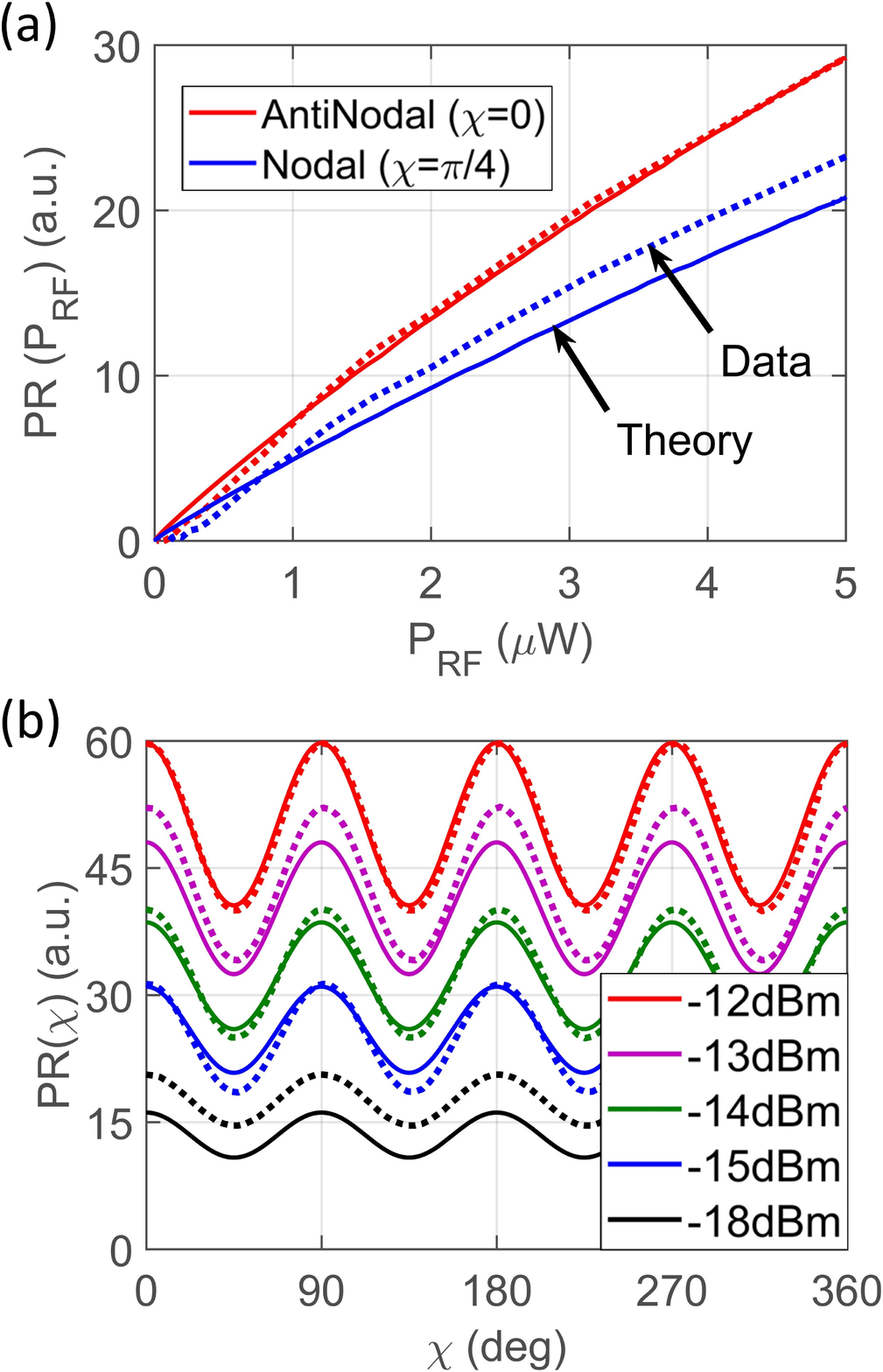}
		\caption{\label{fig:PR_Prf_ang_dep} (a) Input RF power ($P_{RF}$) dependence of total (surface+bulk) PR when $v_s\parallel$ gap antinode ($\chi=0$) and $v_s\parallel$ gap node ($\chi=\pi/4$). The solid lines are the theoretical estimation with the paramagnetic weight factor $p=0.015$ at $T=0.025T_c$ and the dotted lines are the experimental data at $T=3$ K where both temperatures are in the surface response dominant regime. Here, theoretically estimated PR is calculated in arbitrary units. To focus on comparison of the $P_{RF}$ dependence and anisotropy between the antinodal and nodal PR from the theory and experiment, PR from the theory is re-scaled so that the value of the theoretical and experimental PR in the gap antinodal direction at $P_{RF}=5$ $\mu$W are the same. (b) The angular ($\chi$) dependence plot of PR at various $P_{RF}$ shows a 4-fold symmetric pattern which reflects the anisotropic ABS response of the sample. Solid lines are the theoretical estimation curves at $T=0.025T_c$ and dotted lines are fitted curves from the experimental data at $T=4.8$ K from Fig. \ref{fig:expPRangdep}. Again, the same normalization scheme as (a) is used here. PR from the theory is re-scaled so that PR at $\chi=0$, $P_{RF}=-12$ dBm is set to be the same as the experimental value.}
\end{figure}

\par With Eqs. (\ref{PR estimation}),(\ref{deltaW}), the input RF power ($P_{RF}$) dependence and the angular dependence ($\chi$) of the photoresponse at representative $P_{RF}$ is calculated and compared to those from experiment as shown in Fig. \ref{fig:PR_Prf_ang_dep}(a),(b). Here, the thickness of the film $t$ is 300 nm. The zero current penetration depth $\lambda(T)$ which gives temperature dependence in Eq. (\ref{ExpanLambda}) is obtained from $\lambda^2(T)/\lambda^2_0=n/n_s(T)=1/\alpha_{bulk}$\cite{Dahm1997} with $\lambda_0=160$ nm\cite{Tallon1995}. Note that $\alpha_{bulk}$ for the clean limit is used here. The nonlinear coefficients $\beta_{surf}$, $\beta_{bulk}$ (and hence $b_\chi$) are obtained by calculating the third order derivatives of $K_{surf}$, $j_{bulk}$ with respect to $q$:
\begin{equation}\label{betaSurf}
\begin{split}
\beta_{bulk}(T,\chi) &= -\frac{T}{2\pi T_c} \sum_{\omega_n>0}\int^{\pi/2}_{-\pi/2} d\theta \\
& \frac{\Delta^2(4\omega_n^2-\Delta^2)}{(\omega_n^2+\Delta^2)^{7/2}}  \Delta_0^3(0,0)\cos^4\theta  \\
\beta_{surf}(T,\chi) &= -\frac{T}{2\pi T_c} \sum_{\omega_n>0}\int^{\pi/2}_{-\pi/2} d\theta  \\
& \frac{\Delta^2(4\omega_n^4+5\omega_n^2\Delta^2+2\Delta^4)}{\omega_n^4(\omega_n^2+\Delta^2)^{5/2}} \Delta_0^3(0,0)\cos^4\theta 
\end{split}  
\end{equation} 
The modulation in $\beta_{surf}(T), b_\chi(T)$ is estimated by $\delta \beta_{surf} = \partial\beta_{surf}/\partial T\times \delta T$ and $\delta b_\chi=\partial b_\chi/\partial T \times \delta T$. Since $\delta T$ is independent of $P_{RF}$ and $\chi$, it is set to be a proportionality constant. $P_{RF}$ is assumed to be proportional to $q^2$, which is true for the low $P_{RF}$ regime where the external magnetic field does not activate a defect hotspot response.\cite{Kurter2011PRB,Zhuravel2012PRB} This threshold $P_{RF}$ for hotspot activation is $\sim-12$ dBm in our experimental setup as seen from Fig. \ref{fig:expPRqdep}(a). For the spiral sample tested here, the PR crossover temperature $T_{cross}$ where antinodal PR ($\chi=0$) becomes larger than nodal PR ($\chi=\pi/4$) is $\sim 5.6$ K. The $P_{RF}$ and $\chi$ dependence of PR are measured well below this temperature ($T=3$ K, $4.8$ K) where the surface response dominates the total PR. For direct comparison between experiment and theory, PR is theoretically calculated with the choice of the paramagnetic weight factor $p=0.015$ in order to give similar $T_{cross}\sim 0.057 T_c$ as the experimental value, and the $P_{RF}$ and $\chi$ dependence of PR is estimated at about half of the PR crossover temperature $T=0.025T_c \sim T_{cross}/2$ which again ensures the surface PR dominates.    

\par As seen from Fig. \ref{fig:PR_Prf_ang_dep}(a), in the theoretical estimation, PR increases as $P_{RF}$ increases since larger external field drives larger superfluid momentum $q$. Also, antinodal ($\chi=0$) PR is larger than nodal ($\chi=\pi/4$) PR, which is expected for the surface ABS response dominant regime. The anisotropy between antinodal and nodal PR remains about the same throughout the whole $P_{RF}$ range where the PR is estimated. Note that these estimated behaviors of the $P_{RF}$ dependence agree well with those of the experimental data plotted together in Fig. \ref{fig:PR_Prf_ang_dep}(a).

\par As presented in Fig. \ref{fig:PR_Prf_ang_dep}(b), the theoretical angular dependence of PR shows a 4-fold symmetric pattern which is a signature of the ABS anisotropy. Again, the theoretical and experimental angular dependence agree with each other for most of the $P_{RF}$ except for the lowest $P_{RF}$ case (-18 dBm). The minor deviation between experiment and theory is due to the nonlinear response of the microwave detector diode at low $P_{RF}$. The fact that the $P_{RF}$ and angular dependence results from the presented theoretical estimation are in good agreement with the experimental data confirms that the microscopic model is consistent with the measured photoresponse, and especially, is valid to predict the response from surface Andreev bound states under microwave excitation.

\par Throughout this section, the crossover behavior of the surface ABS response and the bulk Meissner response are theoretically described and the $P_{RF}$ and angular dependence of PR is estimated only in terms of the stored energy. As a further extension of this work, it will be important to experimentally measure the $P_{RF}$ dependence of the quality factor $Q$ of the sample and expand the current microscopic theory to understand the loss mechanisms in the ABS. With this detailed understanding, a proper description of the resistive photoresponse can also be obtained.

\section{Conclusions}
\par Making use of the rf resonant technique combined with laser scanning microscopy allows one to visualize the anisotropy of the paramagnetic nonlinear Meissner response from the surface ABS. This image gives crucial information to help determine the gap nodal structure. At low temperature, this gap nodal spectroscopy using ABS response creates a clear anisotropic image for nodal superconductors compared to that arising from the bulk diamagnetic response. A theory correctly describes the observed anisotropy and RF power dependence of the ABS photoresponse.

\begin{acknowledgments}
This work is supported by Volkswagen Foundation grant No. 90284 and work at Maryland is supported by NSF grant No. DMR-1410712 and the Maryland Center for Nanophysics and Advanced Materials. This work is also supported in part by the Ministry of Education and Science of Russian Federation in the framework of Increase Competitiveness Program of the NUST MISIS (contracts no. K2-2014-025, K2-2015-002, and K2-2016-063). S.N.S. acknowledges partial support from the State Fund for Fundamental Research of Ukraine (F66/95-2016). 
\end{acknowledgments}

\bibliography{Ref_ABS_PR}

\begin{thebibliography}{80}%
\makeatletter
\providecommand \@ifxundefined [1]{%
 \@ifx{#1\undefined}
}%
\providecommand \@ifnum [1]{%
 \ifnum #1\expandafter \@firstoftwo
 \else \expandafter \@secondoftwo
 \fi
}%
\providecommand \@ifx [1]{%
 \ifx #1\expandafter \@firstoftwo
 \else \expandafter \@secondoftwo
 \fi
}%
\providecommand \natexlab [1]{#1}%
\providecommand \enquote  [1]{``#1''}%
\providecommand \bibnamefont  [1]{#1}%
\providecommand \bibfnamefont [1]{#1}%
\providecommand \citenamefont [1]{#1}%
\providecommand \href@noop [0]{\@secondoftwo}%
\providecommand \href [0]{\begingroup \@sanitize@url \@href}%
\providecommand \@href[1]{\@@startlink{#1}\@@href}%
\providecommand \@@href[1]{\endgroup#1\@@endlink}%
\providecommand \@sanitize@url [0]{\catcode `\\12\catcode `\$12\catcode
  `\&12\catcode `\#12\catcode `\^12\catcode `\_12\catcode `\%12\relax}%
\providecommand \@@startlink[1]{}%
\providecommand \@@endlink[0]{}%
\providecommand \url  [0]{\begingroup\@sanitize@url \@url }%
\providecommand \@url [1]{\endgroup\@href {#1}{\urlprefix }}%
\providecommand \urlprefix  [0]{URL }%
\providecommand \Eprint [0]{\href }%
\providecommand \doibase [0]{http://dx.doi.org/}%
\providecommand \selectlanguage [0]{\@gobble}%
\providecommand \bibinfo  [0]{\@secondoftwo}%
\providecommand \bibfield  [0]{\@secondoftwo}%
\providecommand \translation [1]{[#1]}%
\providecommand \BibitemOpen [0]{}%
\providecommand \bibitemStop [0]{}%
\providecommand \bibitemNoStop [0]{.\EOS\space}%
\providecommand \EOS [0]{\spacefactor3000\relax}%
\providecommand \BibitemShut  [1]{\csname bibitem#1\endcsname}%
\let\auto@bib@innerbib\@empty
\bibitem [{\citenamefont {Yip}\ and\ \citenamefont {Sauls}(1992)}]{Yip1992}%
  \BibitemOpen
  \bibfield  {author} {\bibinfo {author} {\bibfnamefont {S.~K.}\ \bibnamefont
  {Yip}}\ and\ \bibinfo {author} {\bibfnamefont {J.~A.}\ \bibnamefont
  {Sauls}},\ }\href {\doibase 10.1103/PhysRevLett.69.2264} {\bibfield
  {journal} {\bibinfo  {journal} {Phys. Rev. Lett.}\ }\textbf {\bibinfo
  {volume} {69}},\ \bibinfo {pages} {2264} (\bibinfo {year}
  {1992})}\BibitemShut {NoStop}%
\bibitem [{\citenamefont {Xu}\ \emph {et~al.}(1995)\citenamefont {Xu},
  \citenamefont {Yip},\ and\ \citenamefont {Sauls}}]{DXu1995}%
  \BibitemOpen
  \bibfield  {author} {\bibinfo {author} {\bibfnamefont {D.}~\bibnamefont
  {Xu}}, \bibinfo {author} {\bibfnamefont {S.~K.}\ \bibnamefont {Yip}}, \ and\
  \bibinfo {author} {\bibfnamefont {J.~A.}\ \bibnamefont {Sauls}},\ }\href
  {\doibase 10.1103/PhysRevB.51.16233} {\bibfield  {journal} {\bibinfo
  {journal} {Phys. Rev. B}\ }\textbf {\bibinfo {volume} {51}},\ \bibinfo
  {pages} {16233} (\bibinfo {year} {1995})}\BibitemShut {NoStop}%
\bibitem [{\citenamefont {Groll}\ \emph {et~al.}(2010)\citenamefont {Groll},
  \citenamefont {Gurevich},\ and\ \citenamefont {Chiorescu}}]{Groll2010}%
  \BibitemOpen
  \bibfield  {author} {\bibinfo {author} {\bibfnamefont {N.}~\bibnamefont
  {Groll}}, \bibinfo {author} {\bibfnamefont {A.}~\bibnamefont {Gurevich}}, \
  and\ \bibinfo {author} {\bibfnamefont {I.}~\bibnamefont {Chiorescu}},\ }\href
  {\doibase 10.1103/PhysRevB.81.020504} {\bibfield  {journal} {\bibinfo
  {journal} {Phys. Rev. B}\ }\textbf {\bibinfo {volume} {81}},\ \bibinfo
  {pages} {020504} (\bibinfo {year} {2010})}\BibitemShut {NoStop}%
\bibitem [{\citenamefont {Gittleman}\ \emph {et~al.}(1965)\citenamefont
  {Gittleman}, \citenamefont {Rosenblum}, \citenamefont {Seidel},\ and\
  \citenamefont {Wicklund}}]{Gittleman1965PR}%
  \BibitemOpen
  \bibfield  {author} {\bibinfo {author} {\bibfnamefont {J.}~\bibnamefont
  {Gittleman}}, \bibinfo {author} {\bibfnamefont {B.}~\bibnamefont
  {Rosenblum}}, \bibinfo {author} {\bibfnamefont {T.~E.}\ \bibnamefont
  {Seidel}}, \ and\ \bibinfo {author} {\bibfnamefont {A.~W.}\ \bibnamefont
  {Wicklund}},\ }\href {\doibase 10.1103/PhysRev.137.A527} {\bibfield
  {journal} {\bibinfo  {journal} {Phys. Rev.}\ }\textbf {\bibinfo {volume}
  {137}},\ \bibinfo {pages} {A527} (\bibinfo {year} {1965})}\BibitemShut
  {NoStop}%
\bibitem [{\citenamefont {Dahm}\ and\ \citenamefont
  {Scalapino}(1996)}]{Dahm1996}%
  \BibitemOpen
  \bibfield  {author} {\bibinfo {author} {\bibfnamefont {T.}~\bibnamefont
  {Dahm}}\ and\ \bibinfo {author} {\bibfnamefont {D.~J.}\ \bibnamefont
  {Scalapino}},\ }\href {\doibase 10.1063/1.116960} {\bibfield  {journal}
  {\bibinfo  {journal} {Applied Physics Letters}\ }\textbf {\bibinfo {volume}
  {69}},\ \bibinfo {pages} {4248} (\bibinfo {year} {1996})}\BibitemShut
  {NoStop}%
\bibitem [{\citenamefont {Dahm}\ and\ \citenamefont
  {Scalapino}(1999)}]{Dahm1999}%
  \BibitemOpen
  \bibfield  {author} {\bibinfo {author} {\bibfnamefont {T.}~\bibnamefont
  {Dahm}}\ and\ \bibinfo {author} {\bibfnamefont {D.~J.}\ \bibnamefont
  {Scalapino}},\ }\href {\doibase 10.1103/PhysRevB.60.13125} {\bibfield
  {journal} {\bibinfo  {journal} {Phys. Rev. B}\ }\textbf {\bibinfo {volume}
  {60}},\ \bibinfo {pages} {13125} (\bibinfo {year} {1999})}\BibitemShut
  {NoStop}%
\bibitem [{\citenamefont {Dahm}\ and\ \citenamefont
  {Scalapino}(1997)}]{Dahm1997}%
  \BibitemOpen
  \bibfield  {author} {\bibinfo {author} {\bibfnamefont {T.}~\bibnamefont
  {Dahm}}\ and\ \bibinfo {author} {\bibfnamefont {D.~J.}\ \bibnamefont
  {Scalapino}},\ }\href {http://dx.doi.org/10.1063/1.364056} {\bibfield
  {journal} {\bibinfo  {journal} {Journal of Applied Physics}\ }\textbf
  {\bibinfo {volume} {81}},\ \bibinfo {pages} {2002} (\bibinfo {year}
  {1997})}\BibitemShut {NoStop}%
\bibitem [{\citenamefont {Li}\ \emph {et~al.}(1998)\citenamefont {Li},
  \citenamefont {Hirschfeld},\ and\ \citenamefont {W\"olfle}}]{Li1998}%
  \BibitemOpen
  \bibfield  {author} {\bibinfo {author} {\bibfnamefont {M.-R.}\ \bibnamefont
  {Li}}, \bibinfo {author} {\bibfnamefont {P.~J.}\ \bibnamefont {Hirschfeld}},
  \ and\ \bibinfo {author} {\bibfnamefont {P.}~\bibnamefont {W\"olfle}},\
  }\href {\doibase 10.1103/PhysRevLett.81.5640} {\bibfield  {journal} {\bibinfo
   {journal} {Phys. Rev. Lett.}\ }\textbf {\bibinfo {volume} {81}},\ \bibinfo
  {pages} {5640} (\bibinfo {year} {1998})}\BibitemShut {NoStop}%
\bibitem [{\citenamefont {Bidinosti}\ \emph {et~al.}(1999)\citenamefont
  {Bidinosti}, \citenamefont {Hardy}, \citenamefont {Bonn},\ and\ \citenamefont
  {Liang}}]{Bidinosti1999}%
  \BibitemOpen
  \bibfield  {author} {\bibinfo {author} {\bibfnamefont {C.~P.}\ \bibnamefont
  {Bidinosti}}, \bibinfo {author} {\bibfnamefont {W.~N.}\ \bibnamefont
  {Hardy}}, \bibinfo {author} {\bibfnamefont {D.~A.}\ \bibnamefont {Bonn}}, \
  and\ \bibinfo {author} {\bibfnamefont {R.}~\bibnamefont {Liang}},\ }\href
  {\doibase 10.1103/PhysRevLett.83.3277} {\bibfield  {journal} {\bibinfo
  {journal} {Phys. Rev. Lett.}\ }\textbf {\bibinfo {volume} {83}},\ \bibinfo
  {pages} {3277} (\bibinfo {year} {1999})}\BibitemShut {NoStop}%
\bibitem [{\citenamefont {Bhattacharya}\ \emph {et~al.}(1999)\citenamefont
  {Bhattacharya}, \citenamefont {Zutic}, \citenamefont {Valls}, \citenamefont
  {Goldman}, \citenamefont {Welp},\ and\ \citenamefont
  {Veal}}]{Bhattacharya1999}%
  \BibitemOpen
  \bibfield  {author} {\bibinfo {author} {\bibfnamefont {A.}~\bibnamefont
  {Bhattacharya}}, \bibinfo {author} {\bibfnamefont {I.}~\bibnamefont {Zutic}},
  \bibinfo {author} {\bibfnamefont {O.~T.}\ \bibnamefont {Valls}}, \bibinfo
  {author} {\bibfnamefont {A.~M.}\ \bibnamefont {Goldman}}, \bibinfo {author}
  {\bibfnamefont {U.}~\bibnamefont {Welp}}, \ and\ \bibinfo {author}
  {\bibfnamefont {B.}~\bibnamefont {Veal}},\ }\href {\doibase
  10.1103/PhysRevLett.82.3132} {\bibfield  {journal} {\bibinfo  {journal}
  {Phys. Rev. Lett.}\ }\textbf {\bibinfo {volume} {82}},\ \bibinfo {pages}
  {3132} (\bibinfo {year} {1999})}\BibitemShut {NoStop}%
\bibitem [{\citenamefont {Carrington}\ \emph {et~al.}(2001)\citenamefont
  {Carrington}, \citenamefont {Manzano}, \citenamefont {Prozorov},
  \citenamefont {Giannetta}, \citenamefont {Kameda},\ and\ \citenamefont
  {Tamegai}}]{Carrington2001}%
  \BibitemOpen
  \bibfield  {author} {\bibinfo {author} {\bibfnamefont {A.}~\bibnamefont
  {Carrington}}, \bibinfo {author} {\bibfnamefont {F.}~\bibnamefont {Manzano}},
  \bibinfo {author} {\bibfnamefont {R.}~\bibnamefont {Prozorov}}, \bibinfo
  {author} {\bibfnamefont {R.~W.}\ \bibnamefont {Giannetta}}, \bibinfo {author}
  {\bibfnamefont {N.}~\bibnamefont {Kameda}}, \ and\ \bibinfo {author}
  {\bibfnamefont {T.}~\bibnamefont {Tamegai}},\ }\href {\doibase
  10.1103/PhysRevLett.86.1074} {\bibfield  {journal} {\bibinfo  {journal}
  {Phys. Rev. Lett.}\ }\textbf {\bibinfo {volume} {86}},\ \bibinfo {pages}
  {1074} (\bibinfo {year} {2001})}\BibitemShut {NoStop}%
\bibitem [{\citenamefont {Oates}\ \emph {et~al.}(2004)\citenamefont {Oates},
  \citenamefont {Park},\ and\ \citenamefont {Koren}}]{Oates2004}%
  \BibitemOpen
  \bibfield  {author} {\bibinfo {author} {\bibfnamefont {D.~E.}\ \bibnamefont
  {Oates}}, \bibinfo {author} {\bibfnamefont {S.-H.}\ \bibnamefont {Park}}, \
  and\ \bibinfo {author} {\bibfnamefont {G.}~\bibnamefont {Koren}},\ }\href
  {\doibase 10.1103/PhysRevLett.93.197001} {\bibfield  {journal} {\bibinfo
  {journal} {Phys. Rev. Lett.}\ }\textbf {\bibinfo {volume} {93}},\ \bibinfo
  {pages} {197001} (\bibinfo {year} {2004})}\BibitemShut {NoStop}%
\bibitem [{\citenamefont {Maeda}\ \emph {et~al.}(1995)\citenamefont {Maeda},
  \citenamefont {Iino}, \citenamefont {Hanaguri}, \citenamefont {Motohira},
  \citenamefont {Kishio},\ and\ \citenamefont {Fukase}}]{Maeda1995}%
  \BibitemOpen
  \bibfield  {author} {\bibinfo {author} {\bibfnamefont {A.}~\bibnamefont
  {Maeda}}, \bibinfo {author} {\bibfnamefont {Y.}~\bibnamefont {Iino}},
  \bibinfo {author} {\bibfnamefont {T.}~\bibnamefont {Hanaguri}}, \bibinfo
  {author} {\bibfnamefont {N.}~\bibnamefont {Motohira}}, \bibinfo {author}
  {\bibfnamefont {K.}~\bibnamefont {Kishio}}, \ and\ \bibinfo {author}
  {\bibfnamefont {T.}~\bibnamefont {Fukase}},\ }\href {\doibase
  10.1103/PhysRevLett.74.1202} {\bibfield  {journal} {\bibinfo  {journal}
  {Phys. Rev. Lett.}\ }\textbf {\bibinfo {volume} {74}},\ \bibinfo {pages}
  {1202} (\bibinfo {year} {1995})}\BibitemShut {NoStop}%
\bibitem [{\citenamefont {Carrington}\ \emph {et~al.}(1999)\citenamefont
  {Carrington}, \citenamefont {Giannetta}, \citenamefont {Kim},\ and\
  \citenamefont {Giapintzakis}}]{Carrington1999}%
  \BibitemOpen
  \bibfield  {author} {\bibinfo {author} {\bibfnamefont {A.}~\bibnamefont
  {Carrington}}, \bibinfo {author} {\bibfnamefont {R.~W.}\ \bibnamefont
  {Giannetta}}, \bibinfo {author} {\bibfnamefont {J.~T.}\ \bibnamefont {Kim}},
  \ and\ \bibinfo {author} {\bibfnamefont {J.}~\bibnamefont {Giapintzakis}},\
  }\href {\doibase 10.1103/PhysRevB.59.R14173} {\bibfield  {journal} {\bibinfo
  {journal} {Phys. Rev. B}\ }\textbf {\bibinfo {volume} {59}},\ \bibinfo
  {pages} {R14173} (\bibinfo {year} {1999})}\BibitemShut {NoStop}%
\bibitem [{\citenamefont {Benz}\ \emph {et~al.}(2001)\citenamefont {Benz},
  \citenamefont {W\"unsch}, \citenamefont {Scherer}, \citenamefont {Neuhaus},\
  and\ \citenamefont {Jutzi}}]{Benz2001}%
  \BibitemOpen
  \bibfield  {author} {\bibinfo {author} {\bibfnamefont {G.}~\bibnamefont
  {Benz}}, \bibinfo {author} {\bibfnamefont {S.}~\bibnamefont {W\"unsch}},
  \bibinfo {author} {\bibfnamefont {T.}~\bibnamefont {Scherer}}, \bibinfo
  {author} {\bibfnamefont {M.}~\bibnamefont {Neuhaus}}, \ and\ \bibinfo
  {author} {\bibfnamefont {W.}~\bibnamefont {Jutzi}},\ }\href {\doibase
  http://dx.doi.org/10.1016/S0921-4534(01)00130-7} {\bibfield  {journal}
  {\bibinfo  {journal} {Physica C: Superconductivity}\ }\textbf {\bibinfo
  {volume} {356}},\ \bibinfo {pages} {122 } (\bibinfo {year}
  {2001})}\BibitemShut {NoStop}%
\bibitem [{\citenamefont {Leong}\ \emph {et~al.}(2005)\citenamefont {Leong},
  \citenamefont {Booth},\ and\ \citenamefont {Schima}}]{Leong2005}%
  \BibitemOpen
  \bibfield  {author} {\bibinfo {author} {\bibfnamefont {K.~T.}\ \bibnamefont
  {Leong}}, \bibinfo {author} {\bibfnamefont {J.~C.}\ \bibnamefont {Booth}}, \
  and\ \bibinfo {author} {\bibfnamefont {S.~A.}\ \bibnamefont {Schima}},\
  }\href {\doibase 10.1109/TASC.2005.849371} {\bibfield  {journal} {\bibinfo
  {journal} {IEEE Transactions on Applied Superconductivity}\ }\textbf
  {\bibinfo {volume} {15}},\ \bibinfo {pages} {3608} (\bibinfo {year}
  {2005})}\BibitemShut {NoStop}%
\bibitem [{\citenamefont {Lee}\ and\ \citenamefont
  {Anlage}(2003{\natexlab{a}})}]{SCLee2003APL}%
  \BibitemOpen
  \bibfield  {author} {\bibinfo {author} {\bibfnamefont {S.-C.}\ \bibnamefont
  {Lee}}\ and\ \bibinfo {author} {\bibfnamefont {S.~M.}\ \bibnamefont
  {Anlage}},\ }\href {\doibase 10.1063/1.1561152} {\bibfield  {journal}
  {\bibinfo  {journal} {Applied Physics Letters}\ }\textbf {\bibinfo {volume}
  {82}},\ \bibinfo {pages} {1893} (\bibinfo {year}
  {2003}{\natexlab{a}})}\BibitemShut {NoStop}%
\bibitem [{\citenamefont {Lee}\ and\ \citenamefont
  {Anlage}(2003{\natexlab{b}})}]{SCLee2003IEEE}%
  \BibitemOpen
  \bibfield  {author} {\bibinfo {author} {\bibfnamefont {S.-C.}\ \bibnamefont
  {Lee}}\ and\ \bibinfo {author} {\bibfnamefont {S.~M.}\ \bibnamefont
  {Anlage}},\ }\href {\doibase 10.1109/TASC.2003.812406} {\bibfield  {journal}
  {\bibinfo  {journal} {IEEE Transactions on Applied Superconductivity}\
  }\textbf {\bibinfo {volume} {13}},\ \bibinfo {pages} {3594} (\bibinfo {year}
  {2003}{\natexlab{b}})}\BibitemShut {NoStop}%
\bibitem [{\citenamefont {Lee}\ \emph {et~al.}(2005{\natexlab{a}})\citenamefont
  {Lee}, \citenamefont {Sullivan}, \citenamefont {Ruchti}, \citenamefont
  {Anlage}, \citenamefont {Palmer}, \citenamefont {Maiorov},\ and\
  \citenamefont {Osquiguil}}]{SCLee2005PRB}%
  \BibitemOpen
  \bibfield  {author} {\bibinfo {author} {\bibfnamefont {S.-C.}\ \bibnamefont
  {Lee}}, \bibinfo {author} {\bibfnamefont {M.}~\bibnamefont {Sullivan}},
  \bibinfo {author} {\bibfnamefont {G.~R.}\ \bibnamefont {Ruchti}}, \bibinfo
  {author} {\bibfnamefont {S.~M.}\ \bibnamefont {Anlage}}, \bibinfo {author}
  {\bibfnamefont {B.~S.}\ \bibnamefont {Palmer}}, \bibinfo {author}
  {\bibfnamefont {B.}~\bibnamefont {Maiorov}}, \ and\ \bibinfo {author}
  {\bibfnamefont {E.}~\bibnamefont {Osquiguil}},\ }\href {\doibase
  10.1103/PhysRevB.71.014507} {\bibfield  {journal} {\bibinfo  {journal} {Phys.
  Rev. B}\ }\textbf {\bibinfo {volume} {71}},\ \bibinfo {pages} {014507}
  (\bibinfo {year} {2005}{\natexlab{a}})}\BibitemShut {NoStop}%
\bibitem [{\citenamefont {Lee}\ \emph {et~al.}(2005{\natexlab{b}})\citenamefont
  {Lee}, \citenamefont {Lee},\ and\ \citenamefont {Anlage}}]{SCLee2005PRB2}%
  \BibitemOpen
  \bibfield  {author} {\bibinfo {author} {\bibfnamefont {S.-C.}\ \bibnamefont
  {Lee}}, \bibinfo {author} {\bibfnamefont {S.-Y.}\ \bibnamefont {Lee}}, \ and\
  \bibinfo {author} {\bibfnamefont {S.~M.}\ \bibnamefont {Anlage}},\ }\href
  {\doibase 10.1103/PhysRevB.72.024527} {\bibfield  {journal} {\bibinfo
  {journal} {Phys. Rev. B}\ }\textbf {\bibinfo {volume} {72}},\ \bibinfo
  {pages} {024527} (\bibinfo {year} {2005}{\natexlab{b}})}\BibitemShut
  {NoStop}%
\bibitem [{\citenamefont {Mircea}\ \emph {et~al.}(2009)\citenamefont {Mircea},
  \citenamefont {Xu},\ and\ \citenamefont {Anlage}}]{Mircea2009PRB}%
  \BibitemOpen
  \bibfield  {author} {\bibinfo {author} {\bibfnamefont {D.~I.}\ \bibnamefont
  {Mircea}}, \bibinfo {author} {\bibfnamefont {H.}~\bibnamefont {Xu}}, \ and\
  \bibinfo {author} {\bibfnamefont {S.~M.}\ \bibnamefont {Anlage}},\ }\href
  {\doibase 10.1103/PhysRevB.80.144505} {\bibfield  {journal} {\bibinfo
  {journal} {Phys. Rev. B}\ }\textbf {\bibinfo {volume} {80}},\ \bibinfo
  {pages} {144505} (\bibinfo {year} {2009})}\BibitemShut {NoStop}%
\bibitem [{\citenamefont {Tai}\ \emph {et~al.}(2013)\citenamefont {Tai},
  \citenamefont {Ghamsari},\ and\ \citenamefont {Anlage}}]{Tai2013IEEE}%
  \BibitemOpen
  \bibfield  {author} {\bibinfo {author} {\bibfnamefont {T.}~\bibnamefont
  {Tai}}, \bibinfo {author} {\bibfnamefont {B.~G.}\ \bibnamefont {Ghamsari}}, \
  and\ \bibinfo {author} {\bibfnamefont {S.~M.}\ \bibnamefont {Anlage}},\
  }\href {\doibase 10.1109/TASC.2012.2227651} {\bibfield  {journal} {\bibinfo
  {journal} {IEEE Transactions on Applied Superconductivity}\ }\textbf
  {\bibinfo {volume} {23}},\ \bibinfo {pages} {7100104} (\bibinfo {year}
  {2013})}\BibitemShut {NoStop}%
\bibitem [{\citenamefont {Tai}\ \emph {et~al.}(2014)\citenamefont {Tai},
  \citenamefont {Ghamsari}, \citenamefont {Bieler}, \citenamefont {Tan},
  \citenamefont {Xi},\ and\ \citenamefont {Anlage}}]{Tai2014APL}%
  \BibitemOpen
  \bibfield  {author} {\bibinfo {author} {\bibfnamefont {T.}~\bibnamefont
  {Tai}}, \bibinfo {author} {\bibfnamefont {B.~G.}\ \bibnamefont {Ghamsari}},
  \bibinfo {author} {\bibfnamefont {T.~R.}\ \bibnamefont {Bieler}}, \bibinfo
  {author} {\bibfnamefont {T.}~\bibnamefont {Tan}}, \bibinfo {author}
  {\bibfnamefont {X.~X.}\ \bibnamefont {Xi}}, \ and\ \bibinfo {author}
  {\bibfnamefont {S.~M.}\ \bibnamefont {Anlage}},\ }\href {\doibase
  10.1063/1.4881880} {\bibfield  {journal} {\bibinfo  {journal} {Applied
  Physics Letters}\ }\textbf {\bibinfo {volume} {104}},\ \bibinfo {pages}
  {232603} (\bibinfo {year} {2014})}\BibitemShut {NoStop}%
\bibitem [{\citenamefont {Tai}\ \emph {et~al.}(2015)\citenamefont {Tai},
  \citenamefont {Ghamsari}, \citenamefont {Bieler},\ and\ \citenamefont
  {Anlage}}]{Tai2015PRB}%
  \BibitemOpen
  \bibfield  {author} {\bibinfo {author} {\bibfnamefont {T.}~\bibnamefont
  {Tai}}, \bibinfo {author} {\bibfnamefont {B.~G.}\ \bibnamefont {Ghamsari}},
  \bibinfo {author} {\bibfnamefont {T.}~\bibnamefont {Bieler}}, \ and\ \bibinfo
  {author} {\bibfnamefont {S.~M.}\ \bibnamefont {Anlage}},\ }\href {\doibase
  10.1103/PhysRevB.92.134513} {\bibfield  {journal} {\bibinfo  {journal} {Phys.
  Rev. B}\ }\textbf {\bibinfo {volume} {92}},\ \bibinfo {pages} {134513}
  (\bibinfo {year} {2015})}\BibitemShut {NoStop}%
\bibitem [{\citenamefont {Tai}\ \emph {et~al.}(2017)\citenamefont {Tai},
  \citenamefont {Ghamsari}, \citenamefont {Kang}, \citenamefont {Lee},
  \citenamefont {Eom},\ and\ \citenamefont {Anlage}}]{Tai2017PhysicaC}%
  \BibitemOpen
  \bibfield  {author} {\bibinfo {author} {\bibfnamefont {T.}~\bibnamefont
  {Tai}}, \bibinfo {author} {\bibfnamefont {B.~G.}\ \bibnamefont {Ghamsari}},
  \bibinfo {author} {\bibfnamefont {J.}~\bibnamefont {Kang}}, \bibinfo {author}
  {\bibfnamefont {S.}~\bibnamefont {Lee}}, \bibinfo {author} {\bibfnamefont
  {C.}~\bibnamefont {Eom}}, \ and\ \bibinfo {author} {\bibfnamefont {S.~M.}\
  \bibnamefont {Anlage}},\ }\href {\doibase
  https://doi.org/10.1016/j.physc.2016.11.014} {\bibfield  {journal} {\bibinfo
  {journal} {Physica C: Superconductivity and its Applications}\ }\textbf
  {\bibinfo {volume} {532}},\ \bibinfo {pages} {44 } (\bibinfo {year}
  {2017})}\BibitemShut {NoStop}%
\bibitem [{\citenamefont {Zhuravel}\ \emph {et~al.}(2013)\citenamefont
  {Zhuravel}, \citenamefont {Ghamsari}, \citenamefont {Kurter}, \citenamefont
  {Jung}, \citenamefont {Remillard}, \citenamefont {Abrahams}, \citenamefont
  {Lukashenko}, \citenamefont {Ustinov},\ and\ \citenamefont
  {Anlage}}]{Zhuravel2013}%
  \BibitemOpen
  \bibfield  {author} {\bibinfo {author} {\bibfnamefont {A.~P.}\ \bibnamefont
  {Zhuravel}}, \bibinfo {author} {\bibfnamefont {B.~G.}\ \bibnamefont
  {Ghamsari}}, \bibinfo {author} {\bibfnamefont {C.}~\bibnamefont {Kurter}},
  \bibinfo {author} {\bibfnamefont {P.}~\bibnamefont {Jung}}, \bibinfo {author}
  {\bibfnamefont {S.}~\bibnamefont {Remillard}}, \bibinfo {author}
  {\bibfnamefont {J.}~\bibnamefont {Abrahams}}, \bibinfo {author}
  {\bibfnamefont {A.~V.}\ \bibnamefont {Lukashenko}}, \bibinfo {author}
  {\bibfnamefont {A.~V.}\ \bibnamefont {Ustinov}}, \ and\ \bibinfo {author}
  {\bibfnamefont {S.~M.}\ \bibnamefont {Anlage}},\ }\href {\doibase
  10.1103/PhysRevLett.110.087002} {\bibfield  {journal} {\bibinfo  {journal}
  {Phys. Rev. Lett.}\ }\textbf {\bibinfo {volume} {110}},\ \bibinfo {pages}
  {087002} (\bibinfo {year} {2013})}\BibitemShut {NoStop}%
\bibitem [{\citenamefont {Hu}(1994)}]{Hu1994}%
  \BibitemOpen
  \bibfield  {author} {\bibinfo {author} {\bibfnamefont {C.-R.}\ \bibnamefont
  {Hu}},\ }\href {\doibase 10.1103/PhysRevLett.72.1526} {\bibfield  {journal}
  {\bibinfo  {journal} {Phys. Rev. Lett.}\ }\textbf {\bibinfo {volume} {72}},\
  \bibinfo {pages} {1526} (\bibinfo {year} {1994})}\BibitemShut {NoStop}%
\bibitem [{\citenamefont {Aprili}\ \emph {et~al.}(1999)\citenamefont {Aprili},
  \citenamefont {Badica},\ and\ \citenamefont {Greene}}]{Aprili1999}%
  \BibitemOpen
  \bibfield  {author} {\bibinfo {author} {\bibfnamefont {M.}~\bibnamefont
  {Aprili}}, \bibinfo {author} {\bibfnamefont {E.}~\bibnamefont {Badica}}, \
  and\ \bibinfo {author} {\bibfnamefont {L.~H.}\ \bibnamefont {Greene}},\
  }\href {\doibase 10.1103/PhysRevLett.83.4630} {\bibfield  {journal} {\bibinfo
   {journal} {Phys. Rev. Lett.}\ }\textbf {\bibinfo {volume} {83}},\ \bibinfo
  {pages} {4630} (\bibinfo {year} {1999})}\BibitemShut {NoStop}%
\bibitem [{\citenamefont {Fogelstr\"om}\ \emph {et~al.}(1997)\citenamefont
  {Fogelstr\"om}, \citenamefont {Rainer},\ and\ \citenamefont
  {Sauls}}]{Fogelstrom1997}%
  \BibitemOpen
  \bibfield  {author} {\bibinfo {author} {\bibfnamefont {M.}~\bibnamefont
  {Fogelstr\"om}}, \bibinfo {author} {\bibfnamefont {D.}~\bibnamefont
  {Rainer}}, \ and\ \bibinfo {author} {\bibfnamefont {J.~A.}\ \bibnamefont
  {Sauls}},\ }\href {\doibase 10.1103/PhysRevLett.79.281} {\bibfield  {journal}
  {\bibinfo  {journal} {Phys. Rev. Lett.}\ }\textbf {\bibinfo {volume} {79}},\
  \bibinfo {pages} {281} (\bibinfo {year} {1997})}\BibitemShut {NoStop}%
\bibitem [{\citenamefont {Higashitani}(1997)}]{Higashitani1997}%
  \BibitemOpen
  \bibfield  {author} {\bibinfo {author} {\bibfnamefont {S.}~\bibnamefont
  {Higashitani}},\ }\href {\doibase 10.1143/JPSJ.66.2556} {\bibfield  {journal}
  {\bibinfo  {journal} {Journal of the Physical Society of Japan}\ }\textbf
  {\bibinfo {volume} {66}},\ \bibinfo {pages} {2556} (\bibinfo {year}
  {1997})}\BibitemShut {NoStop}%
\bibitem [{\citenamefont {Braunisch}\ \emph {et~al.}(1992)\citenamefont
  {Braunisch}, \citenamefont {Knauf}, \citenamefont {Kataev}, \citenamefont
  {Neuhausen}, \citenamefont {Gr\"utz}, \citenamefont {Kock}, \citenamefont
  {Roden}, \citenamefont {Khomskii},\ and\ \citenamefont
  {Wohlleben}}]{Braunisch1992}%
  \BibitemOpen
  \bibfield  {author} {\bibinfo {author} {\bibfnamefont {W.}~\bibnamefont
  {Braunisch}}, \bibinfo {author} {\bibfnamefont {N.}~\bibnamefont {Knauf}},
  \bibinfo {author} {\bibfnamefont {V.}~\bibnamefont {Kataev}}, \bibinfo
  {author} {\bibfnamefont {S.}~\bibnamefont {Neuhausen}}, \bibinfo {author}
  {\bibfnamefont {A.}~\bibnamefont {Gr\"utz}}, \bibinfo {author} {\bibfnamefont
  {A.}~\bibnamefont {Kock}}, \bibinfo {author} {\bibfnamefont {B.}~\bibnamefont
  {Roden}}, \bibinfo {author} {\bibfnamefont {D.}~\bibnamefont {Khomskii}}, \
  and\ \bibinfo {author} {\bibfnamefont {D.}~\bibnamefont {Wohlleben}},\ }\href
  {\doibase 10.1103/PhysRevLett.68.1908} {\bibfield  {journal} {\bibinfo
  {journal} {Phys. Rev. Lett.}\ }\textbf {\bibinfo {volume} {68}},\ \bibinfo
  {pages} {1908} (\bibinfo {year} {1992})}\BibitemShut {NoStop}%
\bibitem [{\citenamefont {Walter}\ \emph {et~al.}(1998)\citenamefont {Walter},
  \citenamefont {Prusseit}, \citenamefont {Semerad}, \citenamefont {Kinder},
  \citenamefont {Assmann}, \citenamefont {Huber}, \citenamefont {Burkhardt},
  \citenamefont {Rainer},\ and\ \citenamefont {Sauls}}]{Walter1998PRL}%
  \BibitemOpen
  \bibfield  {author} {\bibinfo {author} {\bibfnamefont {H.}~\bibnamefont
  {Walter}}, \bibinfo {author} {\bibfnamefont {W.}~\bibnamefont {Prusseit}},
  \bibinfo {author} {\bibfnamefont {R.}~\bibnamefont {Semerad}}, \bibinfo
  {author} {\bibfnamefont {H.}~\bibnamefont {Kinder}}, \bibinfo {author}
  {\bibfnamefont {W.}~\bibnamefont {Assmann}}, \bibinfo {author} {\bibfnamefont
  {H.}~\bibnamefont {Huber}}, \bibinfo {author} {\bibfnamefont
  {H.}~\bibnamefont {Burkhardt}}, \bibinfo {author} {\bibfnamefont
  {D.}~\bibnamefont {Rainer}}, \ and\ \bibinfo {author} {\bibfnamefont {J.~A.}\
  \bibnamefont {Sauls}},\ }\href {\doibase 10.1103/PhysRevLett.80.3598}
  {\bibfield  {journal} {\bibinfo  {journal} {Phys. Rev. Lett.}\ }\textbf
  {\bibinfo {volume} {80}},\ \bibinfo {pages} {3598} (\bibinfo {year}
  {1998})}\BibitemShut {NoStop}%
\bibitem [{\citenamefont {Geim}\ \emph {et~al.}(1998)\citenamefont {Geim},
  \citenamefont {Dubonos}, \citenamefont {Lok}, \citenamefont {Henini},\ and\
  \citenamefont {Maan}}]{Geim1998}%
  \BibitemOpen
  \bibfield  {author} {\bibinfo {author} {\bibfnamefont {A.~K.}\ \bibnamefont
  {Geim}}, \bibinfo {author} {\bibfnamefont {S.~V.}\ \bibnamefont {Dubonos}},
  \bibinfo {author} {\bibfnamefont {J.~G.~S.}\ \bibnamefont {Lok}}, \bibinfo
  {author} {\bibfnamefont {M.}~\bibnamefont {Henini}}, \ and\ \bibinfo {author}
  {\bibfnamefont {J.}~\bibnamefont {Maan}},\ }\href
  {http://www.nature.com/nature/journal/v396/n6707/abs/396144a0.html?foxtrotcallback=true#abs}
  {\bibfield  {journal} {\bibinfo  {journal} {Nature}\ }\textbf {\bibinfo
  {volume} {396}},\ \bibinfo {pages} {144} (\bibinfo {year}
  {1998})}\BibitemShut {NoStop}%
\bibitem [{\citenamefont {Barbara}\ \emph {et~al.}(1999)\citenamefont
  {Barbara}, \citenamefont {Araujo-Moreira}, \citenamefont {Cawthorne},\ and\
  \citenamefont {Lobb}}]{Barbara1999PRB}%
  \BibitemOpen
  \bibfield  {author} {\bibinfo {author} {\bibfnamefont {P.}~\bibnamefont
  {Barbara}}, \bibinfo {author} {\bibfnamefont {F.~M.}\ \bibnamefont
  {Araujo-Moreira}}, \bibinfo {author} {\bibfnamefont {A.~B.}\ \bibnamefont
  {Cawthorne}}, \ and\ \bibinfo {author} {\bibfnamefont {C.~J.}\ \bibnamefont
  {Lobb}},\ }\href {\doibase 10.1103/PhysRevB.60.7489} {\bibfield  {journal}
  {\bibinfo  {journal} {Phys. Rev. B}\ }\textbf {\bibinfo {volume} {60}},\
  \bibinfo {pages} {7489} (\bibinfo {year} {1999})}\BibitemShut {NoStop}%
\bibitem [{\citenamefont {Il'ichev}\ \emph {et~al.}(2003)\citenamefont
  {Il'ichev}, \citenamefont {Tafuri}, \citenamefont {Grajcar}, \citenamefont
  {IJsselsteijn}, \citenamefont {Weber}, \citenamefont {Lombardi},\ and\
  \citenamefont {Kirtley}}]{Ilichev2003}%
  \BibitemOpen
  \bibfield  {author} {\bibinfo {author} {\bibfnamefont {E.}~\bibnamefont
  {Il'ichev}}, \bibinfo {author} {\bibfnamefont {F.}~\bibnamefont {Tafuri}},
  \bibinfo {author} {\bibfnamefont {M.}~\bibnamefont {Grajcar}}, \bibinfo
  {author} {\bibfnamefont {R.~P.~J.}\ \bibnamefont {IJsselsteijn}}, \bibinfo
  {author} {\bibfnamefont {J.}~\bibnamefont {Weber}}, \bibinfo {author}
  {\bibfnamefont {F.}~\bibnamefont {Lombardi}}, \ and\ \bibinfo {author}
  {\bibfnamefont {J.~R.}\ \bibnamefont {Kirtley}},\ }\href {\doibase
  10.1103/PhysRevB.68.014510} {\bibfield  {journal} {\bibinfo  {journal} {Phys.
  Rev. B}\ }\textbf {\bibinfo {volume} {68}},\ \bibinfo {pages} {014510}
  (\bibinfo {year} {2003})}\BibitemShut {NoStop}%
\bibitem [{\citenamefont {Li}(2003)}]{Li2003}%
  \BibitemOpen
  \bibfield  {author} {\bibinfo {author} {\bibfnamefont {M.~S.}\ \bibnamefont
  {Li}},\ }\href {\doibase https://doi.org/10.1016/S0370-1573(02)00635-X}
  {\bibfield  {journal} {\bibinfo  {journal} {Physics Reports}\ }\textbf
  {\bibinfo {volume} {376}},\ \bibinfo {pages} {133 } (\bibinfo {year}
  {2003})}\BibitemShut {NoStop}%
\bibitem [{\citenamefont {Barash}\ \emph {et~al.}(2000)\citenamefont {Barash},
  \citenamefont {Kalenkov},\ and\ \citenamefont {Kurkij\"arvi}}]{Barash2000}%
  \BibitemOpen
  \bibfield  {author} {\bibinfo {author} {\bibfnamefont {Y.~S.}\ \bibnamefont
  {Barash}}, \bibinfo {author} {\bibfnamefont {M.~S.}\ \bibnamefont
  {Kalenkov}}, \ and\ \bibinfo {author} {\bibfnamefont {J.}~\bibnamefont
  {Kurkij\"arvi}},\ }\href {\doibase 10.1103/PhysRevB.62.6665} {\bibfield
  {journal} {\bibinfo  {journal} {Phys. Rev. B}\ }\textbf {\bibinfo {volume}
  {62}},\ \bibinfo {pages} {6665} (\bibinfo {year} {2000})}\BibitemShut
  {NoStop}%
\bibitem [{\citenamefont {Zare}\ \emph {et~al.}(2010)\citenamefont {Zare},
  \citenamefont {Dahm},\ and\ \citenamefont {Schopohl}}]{Zare2010PRL}%
  \BibitemOpen
  \bibfield  {author} {\bibinfo {author} {\bibfnamefont {A.}~\bibnamefont
  {Zare}}, \bibinfo {author} {\bibfnamefont {T.}~\bibnamefont {Dahm}}, \ and\
  \bibinfo {author} {\bibfnamefont {N.}~\bibnamefont {Schopohl}},\ }\href
  {\doibase 10.1103/PhysRevLett.104.237001} {\bibfield  {journal} {\bibinfo
  {journal} {Phys. Rev. Lett.}\ }\textbf {\bibinfo {volume} {104}},\ \bibinfo
  {pages} {237001} (\bibinfo {year} {2010})}\BibitemShut {NoStop}%
\bibitem [{\citenamefont {Zhuravel}\ \emph {et~al.}(2012)\citenamefont
  {Zhuravel}, \citenamefont {Kurter}, \citenamefont {Ustinov},\ and\
  \citenamefont {Anlage}}]{Zhuravel2012PRB}%
  \BibitemOpen
  \bibfield  {author} {\bibinfo {author} {\bibfnamefont {A.~P.}\ \bibnamefont
  {Zhuravel}}, \bibinfo {author} {\bibfnamefont {C.}~\bibnamefont {Kurter}},
  \bibinfo {author} {\bibfnamefont {A.~V.}\ \bibnamefont {Ustinov}}, \ and\
  \bibinfo {author} {\bibfnamefont {S.~M.}\ \bibnamefont {Anlage}},\ }\href
  {\doibase 10.1103/PhysRevB.85.134535} {\bibfield  {journal} {\bibinfo
  {journal} {Phys. Rev. B}\ }\textbf {\bibinfo {volume} {85}},\ \bibinfo
  {pages} {134535} (\bibinfo {year} {2012})}\BibitemShut {NoStop}%
\bibitem [{\citenamefont {Ghamsari}\ \emph
  {et~al.}(2013{\natexlab{a}})\citenamefont {Ghamsari}, \citenamefont
  {Abrahams}, \citenamefont {Remillard},\ and\ \citenamefont
  {Anlage}}]{Ghamsari2013APL}%
  \BibitemOpen
  \bibfield  {author} {\bibinfo {author} {\bibfnamefont {B.~G.}\ \bibnamefont
  {Ghamsari}}, \bibinfo {author} {\bibfnamefont {J.}~\bibnamefont {Abrahams}},
  \bibinfo {author} {\bibfnamefont {S.}~\bibnamefont {Remillard}}, \ and\
  \bibinfo {author} {\bibfnamefont {S.~M.}\ \bibnamefont {Anlage}},\ }\href
  {\doibase 10.1063/1.4774080} {\bibfield  {journal} {\bibinfo  {journal}
  {Appl. Phys. Lett}\ }\textbf {\bibinfo {volume} {102}},\ \bibinfo {pages}
  {013503} (\bibinfo {year} {2013}{\natexlab{a}})}\BibitemShut {NoStop}%
\bibitem [{\citenamefont {Ghamsari}\ \emph
  {et~al.}(2013{\natexlab{b}})\citenamefont {Ghamsari}, \citenamefont
  {Abrahams}, \citenamefont {Remillard},\ and\ \citenamefont
  {Anlage}}]{Ghamsari2013IEEE}%
  \BibitemOpen
  \bibfield  {author} {\bibinfo {author} {\bibfnamefont {B.~G.}\ \bibnamefont
  {Ghamsari}}, \bibinfo {author} {\bibfnamefont {J.}~\bibnamefont {Abrahams}},
  \bibinfo {author} {\bibfnamefont {S.}~\bibnamefont {Remillard}}, \ and\
  \bibinfo {author} {\bibfnamefont {S.~M.}\ \bibnamefont {Anlage}},\ }\href
  {\doibase 10.1109/TASC.2012.2232343} {\bibfield  {journal} {\bibinfo
  {journal} {IEEE Transactions on Applied Superconductivity}\ }\textbf
  {\bibinfo {volume} {23}},\ \bibinfo {pages} {1500304} (\bibinfo {year}
  {2013}{\natexlab{b}})}\BibitemShut {NoStop}%
\bibitem [{\citenamefont {Anlage}(2011)}]{Anlage2011JOptics}%
  \BibitemOpen
  \bibfield  {author} {\bibinfo {author} {\bibfnamefont {S.~M.}\ \bibnamefont
  {Anlage}},\ }\href {http://stacks.iop.org/2040-8986/13/i=2/a=024001}
  {\bibfield  {journal} {\bibinfo  {journal} {Journal of Optics}\ }\textbf
  {\bibinfo {volume} {13}},\ \bibinfo {pages} {024001} (\bibinfo {year}
  {2011})}\BibitemShut {NoStop}%
\bibitem [{\citenamefont {Kurter}\ \emph {et~al.}(2010)\citenamefont {Kurter},
  \citenamefont {Abrahams},\ and\ \citenamefont {Anlage}}]{Kurter2010}%
  \BibitemOpen
  \bibfield  {author} {\bibinfo {author} {\bibfnamefont {C.}~\bibnamefont
  {Kurter}}, \bibinfo {author} {\bibfnamefont {J.}~\bibnamefont {Abrahams}}, \
  and\ \bibinfo {author} {\bibfnamefont {S.~M.}\ \bibnamefont {Anlage}},\
  }\href {\doibase 10.1063/1.3456524} {\bibfield  {journal} {\bibinfo
  {journal} {Appl. Phys. Lett}\ }\textbf {\bibinfo {volume} {96}},\ \bibinfo
  {pages} {253504} (\bibinfo {year} {2010})}\BibitemShut {NoStop}%
\bibitem [{\citenamefont {Kurter}\ \emph
  {et~al.}(2011{\natexlab{a}})\citenamefont {Kurter}, \citenamefont {Zhuravel},
  \citenamefont {Ustinov},\ and\ \citenamefont {Anlage}}]{Kurter2011PRB}%
  \BibitemOpen
  \bibfield  {author} {\bibinfo {author} {\bibfnamefont {C.}~\bibnamefont
  {Kurter}}, \bibinfo {author} {\bibfnamefont {A.~P.}\ \bibnamefont
  {Zhuravel}}, \bibinfo {author} {\bibfnamefont {A.~V.}\ \bibnamefont
  {Ustinov}}, \ and\ \bibinfo {author} {\bibfnamefont {S.~M.}\ \bibnamefont
  {Anlage}},\ }\href {\doibase 10.1103/PhysRevB.84.104515} {\bibfield
  {journal} {\bibinfo  {journal} {Phys. Rev. B}\ }\textbf {\bibinfo {volume}
  {84}},\ \bibinfo {pages} {104515} (\bibinfo {year}
  {2011}{\natexlab{a}})}\BibitemShut {NoStop}%
\bibitem [{\citenamefont {Maleeva}\ \emph {et~al.}(2015)\citenamefont
  {Maleeva}, \citenamefont {Averkin}, \citenamefont {Abramov}, \citenamefont
  {Fistul}, \citenamefont {Karpov}, \citenamefont {Zhuravel},\ and\
  \citenamefont {Ustinov}}]{Maleeva2015}%
  \BibitemOpen
  \bibfield  {author} {\bibinfo {author} {\bibfnamefont {N.}~\bibnamefont
  {Maleeva}}, \bibinfo {author} {\bibfnamefont {A.}~\bibnamefont {Averkin}},
  \bibinfo {author} {\bibfnamefont {N.~N.}\ \bibnamefont {Abramov}}, \bibinfo
  {author} {\bibfnamefont {M.~V.}\ \bibnamefont {Fistul}}, \bibinfo {author}
  {\bibfnamefont {A.}~\bibnamefont {Karpov}}, \bibinfo {author} {\bibfnamefont
  {A.~P.}\ \bibnamefont {Zhuravel}}, \ and\ \bibinfo {author} {\bibfnamefont
  {A.~V.}\ \bibnamefont {Ustinov}},\ }\href {\doibase 10.1063/1.4923305}
  {\bibfield  {journal} {\bibinfo  {journal} {J. Appl. Phys}\ }\textbf
  {\bibinfo {volume} {118}},\ \bibinfo {pages} {033902} (\bibinfo {year}
  {2015})}\BibitemShut {NoStop}%
\bibitem [{\citenamefont {Culbertson}\ \emph {et~al.}(1998)\citenamefont
  {Culbertson}, \citenamefont {Newman},\ and\ \citenamefont
  {Wilker}}]{Culbertson1998}%
  \BibitemOpen
  \bibfield  {author} {\bibinfo {author} {\bibfnamefont {J.~C.}\ \bibnamefont
  {Culbertson}}, \bibinfo {author} {\bibfnamefont {H.~S.}\ \bibnamefont
  {Newman}}, \ and\ \bibinfo {author} {\bibfnamefont {C.}~\bibnamefont
  {Wilker}},\ }\href {\doibase 10.1063/1.368390} {\bibfield  {journal}
  {\bibinfo  {journal} {J. Appl. Phys.}\ }\textbf {\bibinfo {volume} {84}},\
  \bibinfo {pages} {2768} (\bibinfo {year} {1998})}\BibitemShut {NoStop}%
\bibitem [{\citenamefont {Kurter}\ \emph
  {et~al.}(2011{\natexlab{b}})\citenamefont {Kurter}, \citenamefont {Zhuravel},
  \citenamefont {Abrahams}, \citenamefont {Bennett}, \citenamefont {Ustinov},\
  and\ \citenamefont {Anlage}}]{Kurter2011IEEE}%
  \BibitemOpen
  \bibfield  {author} {\bibinfo {author} {\bibfnamefont {C.}~\bibnamefont
  {Kurter}}, \bibinfo {author} {\bibfnamefont {A.~P.}\ \bibnamefont
  {Zhuravel}}, \bibinfo {author} {\bibfnamefont {J.}~\bibnamefont {Abrahams}},
  \bibinfo {author} {\bibfnamefont {C.~L.}\ \bibnamefont {Bennett}}, \bibinfo
  {author} {\bibfnamefont {A.~V.}\ \bibnamefont {Ustinov}}, \ and\ \bibinfo
  {author} {\bibfnamefont {S.~M.}\ \bibnamefont {Anlage}},\ }\href {\doibase
  10.1109/TASC.2010.2088093} {\bibfield  {journal} {\bibinfo  {journal} {IEEE
  Transactions on Applied Superconductivity}\ }\textbf {\bibinfo {volume}
  {21}},\ \bibinfo {pages} {709} (\bibinfo {year}
  {2011}{\natexlab{b}})}\BibitemShut {NoStop}%
\bibitem [{\citenamefont {Hooker}\ \emph {et~al.}(2013)\citenamefont {Hooker},
  \citenamefont {Arora}, \citenamefont {Brey}, \citenamefont {Nast},
  \citenamefont {Ramaswamy}, \citenamefont {Edison},\ and\ \citenamefont
  {Withers}}]{Hooker2013}%
  \BibitemOpen
  \bibfield  {author} {\bibinfo {author} {\bibfnamefont {J.~W.}\ \bibnamefont
  {Hooker}}, \bibinfo {author} {\bibfnamefont {R.~K.}\ \bibnamefont {Arora}},
  \bibinfo {author} {\bibfnamefont {W.~W.}\ \bibnamefont {Brey}}, \bibinfo
  {author} {\bibfnamefont {R.~E.}\ \bibnamefont {Nast}}, \bibinfo {author}
  {\bibfnamefont {V.}~\bibnamefont {Ramaswamy}}, \bibinfo {author}
  {\bibfnamefont {A.~S.}\ \bibnamefont {Edison}}, \ and\ \bibinfo {author}
  {\bibfnamefont {R.~S.}\ \bibnamefont {Withers}},\ }in\ \href {\doibase
  10.1109/ISEC.2013.6604295} {\emph {\bibinfo {booktitle} {2013 IEEE 14th
  International Superconductive Electronics Conference (ISEC)}}}\ (\bibinfo
  {year} {2013})\ pp.\ \bibinfo {pages} {1--3}\BibitemShut {NoStop}%
\bibitem [{\citenamefont {Maleeva}\ \emph {et~al.}(2014)\citenamefont
  {Maleeva}, \citenamefont {Fistul}, \citenamefont {Karpov}, \citenamefont
  {Zhuravel}, \citenamefont {Averkin}, \citenamefont {Jung},\ and\
  \citenamefont {Ustinov}}]{Maleeva2014}%
  \BibitemOpen
  \bibfield  {author} {\bibinfo {author} {\bibfnamefont {N.}~\bibnamefont
  {Maleeva}}, \bibinfo {author} {\bibfnamefont {M.~V.}\ \bibnamefont {Fistul}},
  \bibinfo {author} {\bibfnamefont {A.}~\bibnamefont {Karpov}}, \bibinfo
  {author} {\bibfnamefont {A.~P.}\ \bibnamefont {Zhuravel}}, \bibinfo {author}
  {\bibfnamefont {A.}~\bibnamefont {Averkin}}, \bibinfo {author} {\bibfnamefont
  {P.}~\bibnamefont {Jung}}, \ and\ \bibinfo {author} {\bibfnamefont {A.~V.}\
  \bibnamefont {Ustinov}},\ }\href {\doibase 10.1063/1.4863835} {\bibfield
  {journal} {\bibinfo  {journal} {J. Appl Phys}\ }\textbf {\bibinfo {volume}
  {115}},\ \bibinfo {pages} {064910} (\bibinfo {year} {2014})}\BibitemShut
  {NoStop}%
\bibitem [{\citenamefont {Zhuravel}\ \emph
  {et~al.}(2006{\natexlab{a}})\citenamefont {Zhuravel}, \citenamefont
  {Sivakov}, \citenamefont {Turutanov}, \citenamefont {Omelyanchouk},
  \citenamefont {Anlage}, \citenamefont {Lukashenko}, \citenamefont {Ustinov},\
  and\ \citenamefont {Abraimov}}]{Zhuravel2006LTP}%
  \BibitemOpen
  \bibfield  {author} {\bibinfo {author} {\bibfnamefont {A.~P.}\ \bibnamefont
  {Zhuravel}}, \bibinfo {author} {\bibfnamefont {A.~G.}\ \bibnamefont
  {Sivakov}}, \bibinfo {author} {\bibfnamefont {O.~G.}\ \bibnamefont
  {Turutanov}}, \bibinfo {author} {\bibfnamefont {A.~N.}\ \bibnamefont
  {Omelyanchouk}}, \bibinfo {author} {\bibfnamefont {S.~M.}\ \bibnamefont
  {Anlage}}, \bibinfo {author} {\bibfnamefont {A.}~\bibnamefont {Lukashenko}},
  \bibinfo {author} {\bibfnamefont {A.~V.}\ \bibnamefont {Ustinov}}, \ and\
  \bibinfo {author} {\bibfnamefont {D.}~\bibnamefont {Abraimov}},\ }\href
  {\doibase 10.1063/1.2215376} {\bibfield  {journal} {\bibinfo  {journal} {Low
  Temp. Phys}\ }\textbf {\bibinfo {volume} {32}},\ \bibinfo {pages} {592}
  (\bibinfo {year} {2006}{\natexlab{a}})}\BibitemShut {NoStop}%
\bibitem [{\citenamefont {Zhuravel}\ \emph
  {et~al.}(2006{\natexlab{b}})\citenamefont {Zhuravel}, \citenamefont
  {Anlage},\ and\ \citenamefont {Ustinov}}]{Zhuravel2006JSupercond}%
  \BibitemOpen
  \bibfield  {author} {\bibinfo {author} {\bibfnamefont {A.~P.}\ \bibnamefont
  {Zhuravel}}, \bibinfo {author} {\bibfnamefont {S.~M.}\ \bibnamefont
  {Anlage}}, \ and\ \bibinfo {author} {\bibfnamefont {A.~V.}\ \bibnamefont
  {Ustinov}},\ }\href {\doibase 10.1007/s10948-006-0123-5} {\bibfield
  {journal} {\bibinfo  {journal} {J. Supercond. Novel Magnetism}\ }\textbf
  {\bibinfo {volume} {19}},\ \bibinfo {pages} {625} (\bibinfo {year}
  {2006}{\natexlab{b}})}\BibitemShut {NoStop}%
\bibitem [{\citenamefont {Sivakov}\ \emph {et~al.}(1994)\citenamefont
  {Sivakov}, \citenamefont {Zhuravel}, \citenamefont {Turutanov}, \citenamefont
  {Dmitrenko}, \citenamefont {Hilgenkamp}, \citenamefont {Brons}, \citenamefont
  {Flokstra},\ and\ \citenamefont {Rogalla}}]{Sivakov1994}%
  \BibitemOpen
  \bibfield  {author} {\bibinfo {author} {\bibfnamefont {A.}~\bibnamefont
  {Sivakov}}, \bibinfo {author} {\bibfnamefont {A.}~\bibnamefont {Zhuravel}},
  \bibinfo {author} {\bibfnamefont {O.}~\bibnamefont {Turutanov}}, \bibinfo
  {author} {\bibfnamefont {I.}~\bibnamefont {Dmitrenko}}, \bibinfo {author}
  {\bibfnamefont {J.}~\bibnamefont {Hilgenkamp}}, \bibinfo {author}
  {\bibfnamefont {G.}~\bibnamefont {Brons}}, \bibinfo {author} {\bibfnamefont
  {J.}~\bibnamefont {Flokstra}}, \ and\ \bibinfo {author} {\bibfnamefont
  {H.}~\bibnamefont {Rogalla}},\ }\href {\doibase
  http://dx.doi.org/10.1016/0921-4534(94)90298-4} {\bibfield  {journal}
  {\bibinfo  {journal} {Physica C: Superconductivity}\ }\textbf {\bibinfo
  {volume} {232}},\ \bibinfo {pages} {93 } (\bibinfo {year}
  {1994})}\BibitemShut {NoStop}%
\bibitem [{\citenamefont {Kurter}\ \emph {et~al.}(2012)\citenamefont {Kurter},
  \citenamefont {Tassin}, \citenamefont {Zhuravel}, \citenamefont {Zhang},
  \citenamefont {Koschny}, \citenamefont {Ustinov}, \citenamefont {Soukoulis},\
  and\ \citenamefont {Anlage}}]{Kurter2012APL}%
  \BibitemOpen
  \bibfield  {author} {\bibinfo {author} {\bibfnamefont {C.}~\bibnamefont
  {Kurter}}, \bibinfo {author} {\bibfnamefont {P.}~\bibnamefont {Tassin}},
  \bibinfo {author} {\bibfnamefont {A.~P.}\ \bibnamefont {Zhuravel}}, \bibinfo
  {author} {\bibfnamefont {L.}~\bibnamefont {Zhang}}, \bibinfo {author}
  {\bibfnamefont {T.}~\bibnamefont {Koschny}}, \bibinfo {author} {\bibfnamefont
  {A.~V.}\ \bibnamefont {Ustinov}}, \bibinfo {author} {\bibfnamefont {C.~M.}\
  \bibnamefont {Soukoulis}}, \ and\ \bibinfo {author} {\bibfnamefont {S.~M.}\
  \bibnamefont {Anlage}},\ }\href {\doibase 10.1063/1.3696297} {\bibfield
  {journal} {\bibinfo  {journal} {Applied Physics Letters}\ }\textbf {\bibinfo
  {volume} {100}},\ \bibinfo {pages} {121906} (\bibinfo {year}
  {2012})}\BibitemShut {NoStop}%
\bibitem [{\citenamefont {Zhuravel}\ \emph
  {et~al.}(2010{\natexlab{a}})\citenamefont {Zhuravel}, \citenamefont {Anlage},
  \citenamefont {Remillard}, \citenamefont {Lukashenko},\ and\ \citenamefont
  {Ustinov}}]{Zhuravel2010JAP}%
  \BibitemOpen
  \bibfield  {author} {\bibinfo {author} {\bibfnamefont {A.~P.}\ \bibnamefont
  {Zhuravel}}, \bibinfo {author} {\bibfnamefont {S.~M.}\ \bibnamefont
  {Anlage}}, \bibinfo {author} {\bibfnamefont {S.~K.}\ \bibnamefont
  {Remillard}}, \bibinfo {author} {\bibfnamefont {A.~V.}\ \bibnamefont
  {Lukashenko}}, \ and\ \bibinfo {author} {\bibfnamefont {A.~V.}\ \bibnamefont
  {Ustinov}},\ }\href {\doibase 10.1063/1.3467003} {\bibfield  {journal}
  {\bibinfo  {journal} {Journal of Applied Physics}\ }\textbf {\bibinfo
  {volume} {108}},\ \bibinfo {pages} {033920} (\bibinfo {year}
  {2010}{\natexlab{a}})}\BibitemShut {NoStop}%
\bibitem [{\citenamefont {Zhuravel}\ \emph
  {et~al.}(2006{\natexlab{c}})\citenamefont {Zhuravel}, \citenamefont
  {Anlage},\ and\ \citenamefont {Ustinov}}]{Zhuravel2006APL}%
  \BibitemOpen
  \bibfield  {author} {\bibinfo {author} {\bibfnamefont {A.~P.}\ \bibnamefont
  {Zhuravel}}, \bibinfo {author} {\bibfnamefont {S.~M.}\ \bibnamefont
  {Anlage}}, \ and\ \bibinfo {author} {\bibfnamefont {A.~V.}\ \bibnamefont
  {Ustinov}},\ }\href {\doibase 10.1063/1.2205726} {\bibfield  {journal}
  {\bibinfo  {journal} {Applied Physics Letters}\ }\textbf {\bibinfo {volume}
  {88}},\ \bibinfo {pages} {212503} (\bibinfo {year}
  {2006}{\natexlab{c}})}\BibitemShut {NoStop}%
\bibitem [{\citenamefont {Zhuravel}\ \emph
  {et~al.}(2007{\natexlab{a}})\citenamefont {Zhuravel}, \citenamefont {Anlage},
  \citenamefont {Remillard},\ and\ \citenamefont {Ustinov}}]{Zhuravel2007MSMW}%
  \BibitemOpen
  \bibfield  {author} {\bibinfo {author} {\bibfnamefont {A.~P.}\ \bibnamefont
  {Zhuravel}}, \bibinfo {author} {\bibfnamefont {S.~M.}\ \bibnamefont
  {Anlage}}, \bibinfo {author} {\bibfnamefont {S.}~\bibnamefont {Remillard}}, \
  and\ \bibinfo {author} {\bibfnamefont {A.}~\bibnamefont {Ustinov}},\ }in\
  \href {\doibase 10.1109/MSMW.2007.4294677} {\emph {\bibinfo {booktitle}
  {Proceedings of the Sixth International Symposium on Physics and Engineering
  of Microwaves, Millimeter and Sub-millimeter Waves}}},\ Vol.~\bibinfo
  {volume} {1}\ (\bibinfo {year} {2007})\ pp.\ \bibinfo {pages}
  {404--406}\BibitemShut {NoStop}%
\bibitem [{\citenamefont {Zhuravel}\ \emph
  {et~al.}(2010{\natexlab{b}})\citenamefont {Zhuravel}, \citenamefont
  {Anlage},\ and\ \citenamefont {Ustinov}}]{Zhuravel2010MSMW}%
  \BibitemOpen
  \bibfield  {author} {\bibinfo {author} {\bibfnamefont {A.~P.}\ \bibnamefont
  {Zhuravel}}, \bibinfo {author} {\bibfnamefont {S.~M.}\ \bibnamefont
  {Anlage}}, \ and\ \bibinfo {author} {\bibfnamefont {A.}~\bibnamefont
  {Ustinov}},\ }in\ \href {\doibase 10.1109/MSMW.2010.5546182} {\emph {\bibinfo
  {booktitle} {Proceedings of the Seventh International Symposium on Physics
  and Engineering of Microwaves, Millimeter and Sub-millimeter Waves}}},\
  Vol.~\bibinfo {volume} {1}\ (\bibinfo {year} {2010})\ p.~\bibinfo {pages}
  {13}\BibitemShut {NoStop}%
\bibitem [{\citenamefont {Zhuravel}\ \emph
  {et~al.}(2007{\natexlab{b}})\citenamefont {Zhuravel}, \citenamefont
  {Anlage},\ and\ \citenamefont {Ustinov}}]{Zhuravel2007IEEE}%
  \BibitemOpen
  \bibfield  {author} {\bibinfo {author} {\bibfnamefont {A.~P.}\ \bibnamefont
  {Zhuravel}}, \bibinfo {author} {\bibfnamefont {S.~M.}\ \bibnamefont
  {Anlage}}, \ and\ \bibinfo {author} {\bibfnamefont {A.~V.}\ \bibnamefont
  {Ustinov}},\ }\href {\doibase 10.1109/TASC.2007.897322} {\bibfield  {journal}
  {\bibinfo  {journal} {IEEE Transactions on Applied Superconductivity}\
  }\textbf {\bibinfo {volume} {17}},\ \bibinfo {pages} {902} (\bibinfo {year}
  {2007}{\natexlab{b}})}\BibitemShut {NoStop}%
\bibitem [{\citenamefont {Zhuravel}\ \emph {et~al.}(2002)\citenamefont
  {Zhuravel}, \citenamefont {Ustinov}, \citenamefont {Harshavardhan},\ and\
  \citenamefont {Anlage}}]{Zhuravel2002APL}%
  \BibitemOpen
  \bibfield  {author} {\bibinfo {author} {\bibfnamefont {A.~P.}\ \bibnamefont
  {Zhuravel}}, \bibinfo {author} {\bibfnamefont {A.~V.}\ \bibnamefont
  {Ustinov}}, \bibinfo {author} {\bibfnamefont {K.~S.}\ \bibnamefont
  {Harshavardhan}}, \ and\ \bibinfo {author} {\bibfnamefont {S.~M.}\
  \bibnamefont {Anlage}},\ }\href {\doibase 10.1063/1.1530753} {\bibfield
  {journal} {\bibinfo  {journal} {Applied Physics Letters}\ }\textbf {\bibinfo
  {volume} {81}},\ \bibinfo {pages} {4979} (\bibinfo {year}
  {2002})}\BibitemShut {NoStop}%
\bibitem [{\citenamefont {Streiffer}\ \emph {et~al.}(1991)\citenamefont
  {Streiffer}, \citenamefont {Zielinski}, \citenamefont {Lairson},\ and\
  \citenamefont {Bravman}}]{Streiffer1991APL}%
  \BibitemOpen
  \bibfield  {author} {\bibinfo {author} {\bibfnamefont {S.~K.}\ \bibnamefont
  {Streiffer}}, \bibinfo {author} {\bibfnamefont {E.~M.}\ \bibnamefont
  {Zielinski}}, \bibinfo {author} {\bibfnamefont {B.~M.}\ \bibnamefont
  {Lairson}}, \ and\ \bibinfo {author} {\bibfnamefont {J.~C.}\ \bibnamefont
  {Bravman}},\ }\href {\doibase 10.1063/1.104996} {\bibfield  {journal}
  {\bibinfo  {journal} {Applied Physics Letters}\ }\textbf {\bibinfo {volume}
  {58}},\ \bibinfo {pages} {2171} (\bibinfo {year} {1991})}\BibitemShut
  {NoStop}%
\bibitem [{\citenamefont {Prozorov}\ and\ \citenamefont
  {Giannetta}(2006{\natexlab{a}})}]{Prozorov2006SUST}%
  \BibitemOpen
  \bibfield  {author} {\bibinfo {author} {\bibfnamefont {R.}~\bibnamefont
  {Prozorov}}\ and\ \bibinfo {author} {\bibfnamefont {R.~W.}\ \bibnamefont
  {Giannetta}},\ }\href {http://stacks.iop.org/0953-2048/19/i=8/a=R01}
  {\bibfield  {journal} {\bibinfo  {journal} {Superconductor Science and
  Technology}\ }\textbf {\bibinfo {volume} {19}},\ \bibinfo {pages} {R41}
  (\bibinfo {year} {2006}{\natexlab{a}})}\BibitemShut {NoStop}%
\bibitem [{\citenamefont {Anlage}\ \emph {et~al.}(1989)\citenamefont {Anlage},
  \citenamefont {Sze}, \citenamefont {Snortland}, \citenamefont {Tahara},
  \citenamefont {Langley}, \citenamefont {Eom}, \citenamefont {Beasley},\ and\
  \citenamefont {Taber}}]{Anlage1989APL}%
  \BibitemOpen
  \bibfield  {author} {\bibinfo {author} {\bibfnamefont {S.~M.}\ \bibnamefont
  {Anlage}}, \bibinfo {author} {\bibfnamefont {H.}~\bibnamefont {Sze}},
  \bibinfo {author} {\bibfnamefont {H.~J.}\ \bibnamefont {Snortland}}, \bibinfo
  {author} {\bibfnamefont {S.}~\bibnamefont {Tahara}}, \bibinfo {author}
  {\bibfnamefont {B.}~\bibnamefont {Langley}}, \bibinfo {author} {\bibfnamefont
  {C.}~\bibnamefont {Eom}}, \bibinfo {author} {\bibfnamefont {M.~R.}\
  \bibnamefont {Beasley}}, \ and\ \bibinfo {author} {\bibfnamefont
  {R.}~\bibnamefont {Taber}},\ }\href {\doibase 10.1063/1.101547} {\bibfield
  {journal} {\bibinfo  {journal} {Applied Physics Letters}\ }\textbf {\bibinfo
  {volume} {54}},\ \bibinfo {pages} {2710} (\bibinfo {year}
  {1989})}\BibitemShut {NoStop}%
\bibitem [{\citenamefont {Talanov}\ \emph {et~al.}(2000)\citenamefont
  {Talanov}, \citenamefont {Mercaldo}, \citenamefont {Anlage},\ and\
  \citenamefont {Claassen}}]{Talanov2000RSI}%
  \BibitemOpen
  \bibfield  {author} {\bibinfo {author} {\bibfnamefont {V.~V.}\ \bibnamefont
  {Talanov}}, \bibinfo {author} {\bibfnamefont {L.~V.}\ \bibnamefont
  {Mercaldo}}, \bibinfo {author} {\bibfnamefont {S.~M.}\ \bibnamefont
  {Anlage}}, \ and\ \bibinfo {author} {\bibfnamefont {J.~H.}\ \bibnamefont
  {Claassen}},\ }\href {\doibase 10.1063/1.1150596} {\bibfield  {journal}
  {\bibinfo  {journal} {Review of Scientific Instruments}\ }\textbf {\bibinfo
  {volume} {71}},\ \bibinfo {pages} {2136} (\bibinfo {year}
  {2000})}\BibitemShut {NoStop}%
\bibitem [{\citenamefont {Mercaldo}\ \emph {et~al.}(2000)\citenamefont
  {Mercaldo}, \citenamefont {Talanov}, \citenamefont {Anlage}, \citenamefont
  {Attanasio},\ and\ \citenamefont {Maritato}}]{Mercaldo2000}%
  \BibitemOpen
  \bibfield  {author} {\bibinfo {author} {\bibfnamefont {L.~V.}\ \bibnamefont
  {Mercaldo}}, \bibinfo {author} {\bibfnamefont {V.~V.}\ \bibnamefont
  {Talanov}}, \bibinfo {author} {\bibfnamefont {S.~M.}\ \bibnamefont {Anlage}},
  \bibinfo {author} {\bibfnamefont {C.}~\bibnamefont {Attanasio}}, \ and\
  \bibinfo {author} {\bibfnamefont {L.}~\bibnamefont {Maritato}},\ }\href
  {\doibase 10.1142/S0217979200003101} {\bibfield  {journal} {\bibinfo
  {journal} {International Journal of Modern Physics B}\ }\textbf {\bibinfo
  {volume} {14}},\ \bibinfo {pages} {2920} (\bibinfo {year}
  {2000})}\BibitemShut {NoStop}%
\bibitem [{\citenamefont {Prozorov}\ and\ \citenamefont
  {Giannetta}(2006{\natexlab{b}})}]{Prozorov2006SST}%
  \BibitemOpen
  \bibfield  {author} {\bibinfo {author} {\bibfnamefont {R.}~\bibnamefont
  {Prozorov}}\ and\ \bibinfo {author} {\bibfnamefont {R.~W.}\ \bibnamefont
  {Giannetta}},\ }\href {http://stacks.iop.org/0953-2048/19/i=8/a=R01}
  {\bibfield  {journal} {\bibinfo  {journal} {Superconductor Science and
  Technology}\ }\textbf {\bibinfo {volume} {19}},\ \bibinfo {pages} {R41}
  (\bibinfo {year} {2006}{\natexlab{b}})}\BibitemShut {NoStop}%
\bibitem [{\citenamefont {Hein}\ \emph {et~al.}(2002)\citenamefont {Hein},
  \citenamefont {Oates}, \citenamefont {Hirst}, \citenamefont {Humphreys},\
  and\ \citenamefont {Velichko}}]{Hein2002APL}%
  \BibitemOpen
  \bibfield  {author} {\bibinfo {author} {\bibfnamefont {M.~A.}\ \bibnamefont
  {Hein}}, \bibinfo {author} {\bibfnamefont {D.~E.}\ \bibnamefont {Oates}},
  \bibinfo {author} {\bibfnamefont {P.~J.}\ \bibnamefont {Hirst}}, \bibinfo
  {author} {\bibfnamefont {R.~G.}\ \bibnamefont {Humphreys}}, \ and\ \bibinfo
  {author} {\bibfnamefont {A.~V.}\ \bibnamefont {Velichko}},\ }\href {\doibase
  10.1063/1.1447000} {\bibfield  {journal} {\bibinfo  {journal} {Applied
  Physics Letters}\ }\textbf {\bibinfo {volume} {80}},\ \bibinfo {pages} {1007}
  (\bibinfo {year} {2002})}\BibitemShut {NoStop}%
\bibitem [{\citenamefont {O\textsc{\char13}Connell}\ \emph
  {et~al.}(2008)\citenamefont {O\textsc{\char13}Connell}, \citenamefont
  {Ansmann}, \citenamefont {Bialczak}, \citenamefont {Hofheinz}, \citenamefont
  {Katz}, \citenamefont {Lucero}, \citenamefont {McKenney}, \citenamefont
  {Neeley}, \citenamefont {Wang}, \citenamefont {Weig}, \citenamefont
  {Cleland},\ and\ \citenamefont {Martinis}}]{O’Connell2008APL}%
  \BibitemOpen
  \bibfield  {author} {\bibinfo {author} {\bibfnamefont {A.~D.}\ \bibnamefont
  {O\textsc{\char13}Connell}}, \bibinfo {author} {\bibfnamefont
  {M.}~\bibnamefont {Ansmann}}, \bibinfo {author} {\bibfnamefont {R.~C.}\
  \bibnamefont {Bialczak}}, \bibinfo {author} {\bibfnamefont {M.}~\bibnamefont
  {Hofheinz}}, \bibinfo {author} {\bibfnamefont {N.}~\bibnamefont {Katz}},
  \bibinfo {author} {\bibfnamefont {E.}~\bibnamefont {Lucero}}, \bibinfo
  {author} {\bibfnamefont {C.}~\bibnamefont {McKenney}}, \bibinfo {author}
  {\bibfnamefont {M.}~\bibnamefont {Neeley}}, \bibinfo {author} {\bibfnamefont
  {H.}~\bibnamefont {Wang}}, \bibinfo {author} {\bibfnamefont {E.~M.}\
  \bibnamefont {Weig}}, \bibinfo {author} {\bibfnamefont {A.~N.}\ \bibnamefont
  {Cleland}}, \ and\ \bibinfo {author} {\bibfnamefont {J.~M.}\ \bibnamefont
  {Martinis}},\ }\href {\doibase 10.1063/1.2898887} {\bibfield  {journal}
  {\bibinfo  {journal} {Applied Physics Letters}\ }\textbf {\bibinfo {volume}
  {92}},\ \bibinfo {pages} {112903} (\bibinfo {year} {2008})}\BibitemShut
  {NoStop}%
\bibitem [{\citenamefont {Pin-Jia}\ \emph {et~al.}(2014)\citenamefont
  {Pin-Jia}, \citenamefont {Yi-Wen},\ and\ \citenamefont
  {Lian-Fu}}]{ZhouPinJia2014CPL}%
  \BibitemOpen
  \bibfield  {author} {\bibinfo {author} {\bibfnamefont {Z.}~\bibnamefont
  {Pin-Jia}}, \bibinfo {author} {\bibfnamefont {W.}~\bibnamefont {Yi-Wen}}, \
  and\ \bibinfo {author} {\bibfnamefont {W.}~\bibnamefont {Lian-Fu}},\ }\href
  {http://stacks.iop.org/0256-307X/31/i=6/a=067402} {\bibfield  {journal}
  {\bibinfo  {journal} {Chin. Phys. Lett.}\ }\textbf {\bibinfo {volume} {31}},\
  \bibinfo {pages} {067402} (\bibinfo {year} {2014})}\BibitemShut {NoStop}%
\bibitem [{\citenamefont {Eilenberger}(1968)}]{Eilenberger1968}%
  \BibitemOpen
  \bibfield  {author} {\bibinfo {author} {\bibfnamefont {G.}~\bibnamefont
  {Eilenberger}},\ }\href
  {https://link.springer.com/article/10.1007/BF01379803} {\bibfield  {journal}
  {\bibinfo  {journal} {Z. Phys.}\ }\textbf {\bibinfo {volume} {214}},\
  \bibinfo {pages} {195} (\bibinfo {year} {1968})}\BibitemShut {NoStop}%
\bibitem [{\citenamefont {Kulik}\ and\ \citenamefont
  {Omelyanchouk}(1978)}]{Kulik1978}%
  \BibitemOpen
  \bibfield  {author} {\bibinfo {author} {\bibfnamefont {I.~O.}\ \bibnamefont
  {Kulik}}\ and\ \bibinfo {author} {\bibfnamefont {A.~N.}\ \bibnamefont
  {Omelyanchouk}},\ }\href@noop {} {\bibfield  {journal} {\bibinfo  {journal}
  {Sov. J. Low Temp. Phys.}\ }\textbf {\bibinfo {volume} {4}},\ \bibinfo
  {pages} {142} (\bibinfo {year} {1978})}\BibitemShut {NoStop}%
\bibitem [{\citenamefont {Belzig}\ \emph {et~al.}(1999)\citenamefont {Belzig},
  \citenamefont {Wilhelm}, \citenamefont {Bruder}, \citenamefont {Sch{\"o}n},\
  and\ \citenamefont {Zaikin}}]{Belzig1999}%
  \BibitemOpen
  \bibfield  {author} {\bibinfo {author} {\bibfnamefont {W.}~\bibnamefont
  {Belzig}}, \bibinfo {author} {\bibfnamefont {F.~K.}\ \bibnamefont {Wilhelm}},
  \bibinfo {author} {\bibfnamefont {C.}~\bibnamefont {Bruder}}, \bibinfo
  {author} {\bibfnamefont {G.}~\bibnamefont {Sch{\"o}n}}, \ and\ \bibinfo
  {author} {\bibfnamefont {A.~D.}\ \bibnamefont {Zaikin}},\ }\href {\doibase
  http://dx.doi.org/10.1006/spmi.1999.0710} {\bibfield  {journal} {\bibinfo
  {journal} {Superlattices and Microstructures}\ }\textbf {\bibinfo {volume}
  {25}},\ \bibinfo {pages} {1251 } (\bibinfo {year} {1999})}\BibitemShut
  {NoStop}%
\bibitem [{\citenamefont {Agassi}\ \emph {et~al.}(2012)\citenamefont {Agassi},
  \citenamefont {Oates},\ and\ \citenamefont {Moeckly}}]{Agassi2012PhysicaC}%
  \BibitemOpen
  \bibfield  {author} {\bibinfo {author} {\bibfnamefont {Y.}~\bibnamefont
  {Agassi}}, \bibinfo {author} {\bibfnamefont {D.}~\bibnamefont {Oates}}, \
  and\ \bibinfo {author} {\bibfnamefont {B.}~\bibnamefont {Moeckly}},\ }\href
  {\doibase https://doi.org/10.1016/j.physc.2012.04.034} {\bibfield  {journal}
  {\bibinfo  {journal} {Physica C: Superconductivity}\ }\textbf {\bibinfo
  {volume} {480}},\ \bibinfo {pages} {79 } (\bibinfo {year}
  {2012})}\BibitemShut {NoStop}%
\bibitem [{\citenamefont {Kolesnichenko}\ \emph {et~al.}(2004)\citenamefont
  {Kolesnichenko}, \citenamefont {Omelyanchouk},\ and\ \citenamefont
  {Shevchenko}}]{Kolesnichenko2004}%
  \BibitemOpen
  \bibfield  {author} {\bibinfo {author} {\bibfnamefont {Y.~A.}\ \bibnamefont
  {Kolesnichenko}}, \bibinfo {author} {\bibfnamefont {A.~N.}\ \bibnamefont
  {Omelyanchouk}}, \ and\ \bibinfo {author} {\bibfnamefont {S.~N.}\
  \bibnamefont {Shevchenko}},\ }\href {\doibase 10.1063/1.1645180} {\bibfield
  {journal} {\bibinfo  {journal} {Low Temp. Phys}\ }\textbf {\bibinfo {volume}
  {30}},\ \bibinfo {pages} {213} (\bibinfo {year} {2004})}\BibitemShut
  {NoStop}%
\bibitem [{\citenamefont {Kolesnichenko}\ \emph {et~al.}(2003)\citenamefont
  {Kolesnichenko}, \citenamefont {Omelyanchouk},\ and\ \citenamefont
  {Shevchenko}}]{Kolesnichenko2003PRB}%
  \BibitemOpen
  \bibfield  {author} {\bibinfo {author} {\bibfnamefont {Y.~A.}\ \bibnamefont
  {Kolesnichenko}}, \bibinfo {author} {\bibfnamefont {A.~N.}\ \bibnamefont
  {Omelyanchouk}}, \ and\ \bibinfo {author} {\bibfnamefont {S.~N.}\
  \bibnamefont {Shevchenko}},\ }\href {\doibase 10.1103/PhysRevB.67.172504}
  {\bibfield  {journal} {\bibinfo  {journal} {Phys. Rev. B}\ }\textbf {\bibinfo
  {volume} {67}},\ \bibinfo {pages} {172504} (\bibinfo {year}
  {2003})}\BibitemShut {NoStop}%
\bibitem [{\citenamefont {Shevchenko}(2009)}]{Shevchenko2009}%
  \BibitemOpen
  \bibfield  {author} {\bibinfo {author} {\bibfnamefont {S.~N.}\ \bibnamefont
  {Shevchenko}},\ }\href {\doibase 10.1063/1.3266914} {\bibfield  {journal}
  {\bibinfo  {journal} {Low Temperature Physics}\ }\textbf {\bibinfo {volume}
  {35}},\ \bibinfo {pages} {854} (\bibinfo {year} {2009})}\BibitemShut
  {NoStop}%
\bibitem [{\citenamefont {Shevchenko}(2006)}]{Shevchenko2006PRB}%
  \BibitemOpen
  \bibfield  {author} {\bibinfo {author} {\bibfnamefont {S.~N.}\ \bibnamefont
  {Shevchenko}},\ }\href {\doibase 10.1103/PhysRevB.74.172502} {\bibfield
  {journal} {\bibinfo  {journal} {Phys. Rev. B}\ }\textbf {\bibinfo {volume}
  {74}},\ \bibinfo {pages} {172502} (\bibinfo {year} {2006})}\BibitemShut
  {NoStop}%
\bibitem [{\citenamefont {Ferrer}\ \emph {et~al.}(1999)\citenamefont {Ferrer},
  \citenamefont {Gonz{\`a}lez-Alvarez},\ and\ \citenamefont
  {S{\`a}nchez-Ca{\~n}izares}}]{Ferrer1999}%
  \BibitemOpen
  \bibfield  {author} {\bibinfo {author} {\bibfnamefont {J.}~\bibnamefont
  {Ferrer}}, \bibinfo {author} {\bibfnamefont {M.}~\bibnamefont
  {Gonz{\`a}lez-Alvarez}}, \ and\ \bibinfo {author} {\bibfnamefont
  {J.}~\bibnamefont {S{\`a}nchez-Ca{\~n}izares}},\ }\href {\doibase
  http://dx.doi.org/10.1006/spmi.1999.0751} {\bibfield  {journal} {\bibinfo
  {journal} {Superlattices and Microstructures}\ }\textbf {\bibinfo {volume}
  {25}},\ \bibinfo {pages} {1125 } (\bibinfo {year} {1999})}\BibitemShut
  {NoStop}%
\bibitem [{\citenamefont {Nicol}\ and\ \citenamefont
  {Carbotte}(2006)}]{EJNicol2006PRB}%
  \BibitemOpen
  \bibfield  {author} {\bibinfo {author} {\bibfnamefont {E.~J.}\ \bibnamefont
  {Nicol}}\ and\ \bibinfo {author} {\bibfnamefont {J.~P.}\ \bibnamefont
  {Carbotte}},\ }\href {\doibase 10.1103/PhysRevB.73.174510} {\bibfield
  {journal} {\bibinfo  {journal} {Phys. Rev. B}\ }\textbf {\bibinfo {volume}
  {73}},\ \bibinfo {pages} {174510} (\bibinfo {year} {2006})}\BibitemShut
  {NoStop}%
\bibitem [{\citenamefont {Rhoderick}\ and\ \citenamefont
  {Wilson}(1962)}]{Rhoderick1962}%
  \BibitemOpen
  \bibfield  {author} {\bibinfo {author} {\bibfnamefont {E.~H.}\ \bibnamefont
  {Rhoderick}}\ and\ \bibinfo {author} {\bibfnamefont {E.~M.}\ \bibnamefont
  {Wilson}},\ }\href {\doibase 10.1038/1941167b0} {\bibfield  {journal}
  {\bibinfo  {journal} {Nature}\ }\textbf {\bibinfo {volume} {194}},\ \bibinfo
  {pages} {1167} (\bibinfo {year} {1962})}\BibitemShut {NoStop}%
\bibitem [{\citenamefont {Tallon}\ \emph {et~al.}(1995)\citenamefont {Tallon},
  \citenamefont {Bernhard}, \citenamefont {Binninger}, \citenamefont {Hofer},
  \citenamefont {Williams}, \citenamefont {Ansaldo}, \citenamefont {Budnick},\
  and\ \citenamefont {Niedermayer}}]{Tallon1995}%
  \BibitemOpen
  \bibfield  {author} {\bibinfo {author} {\bibfnamefont {J.~L.}\ \bibnamefont
  {Tallon}}, \bibinfo {author} {\bibfnamefont {C.}~\bibnamefont {Bernhard}},
  \bibinfo {author} {\bibfnamefont {U.}~\bibnamefont {Binninger}}, \bibinfo
  {author} {\bibfnamefont {A.}~\bibnamefont {Hofer}}, \bibinfo {author}
  {\bibfnamefont {G.~V.~M.}\ \bibnamefont {Williams}}, \bibinfo {author}
  {\bibfnamefont {E.~J.}\ \bibnamefont {Ansaldo}}, \bibinfo {author}
  {\bibfnamefont {J.~I.}\ \bibnamefont {Budnick}}, \ and\ \bibinfo {author}
  {\bibfnamefont {C.}~\bibnamefont {Niedermayer}},\ }\href {\doibase
  10.1103/PhysRevLett.74.1008} {\bibfield  {journal} {\bibinfo  {journal}
  {Phys. Rev. Lett.}\ }\textbf {\bibinfo {volume} {74}},\ \bibinfo {pages}
  {1008} (\bibinfo {year} {1995})}\BibitemShut {NoStop}%
\end{thebibliography}%

\end{document}